\documentclass[twocolumn,floatfix,superscriptaddress]{revtex4}
\usepackage{graphicx}
\usepackage{amsmath}
\usepackage{amssymb}
\usepackage{bm}

\usepackage{color}

\begin{document}

\title{Bimodal conductance distribution of Kitaev edge modes in topological superconductors}
\author{M. Diez}
\affiliation{Instituut-Lorentz, Universiteit Leiden, P.O. Box 9506, 2300 RA Leiden, The Netherlands}
\author{I. C. Fulga}
\affiliation{Instituut-Lorentz, Universiteit Leiden, P.O. Box 9506, 2300 RA Leiden, The Netherlands}
\author{D. I. Pikulin}
\affiliation{Instituut-Lorentz, Universiteit Leiden, P.O. Box 9506, 2300 RA Leiden, The Netherlands}
\author{J. Tworzyd{\l}o} 
\affiliation{Institute of Theoretical Physics, Faculty of Physics, University of Warsaw, Ho\.{z}a 69, 00--681 Warsaw, Poland}
\author{C. W. J. Beenakker}
\affiliation{Instituut-Lorentz, Universiteit Leiden, P.O. Box 9506, 2300 RA Leiden, The Netherlands}
\date{March 2014}
\begin{abstract}
A two-dimensional superconductor with spin-triplet \textit{p}-wave pairing supports chiral or helical Majorana edge modes with a quantized (length $L$-independent) thermal conductance. Sufficiently strong anisotropy removes both chirality and helicity, doubling the conductance in the clean system and imposing a super-Ohmic $1/\sqrt{L}$ decay in the presence of disorder. We explain the absence of localization in the framework of the Kitaev Hamiltonian,  contrasting the edge modes of the two-dimensional system with the one-dimensional Kitaev chain. While the disordered Kitaev chain has a log-normal conductance distribution peaked at an exponentially small value, the Kitaev edge has a bimodal distribution with a second peak near the conductance quantum. Shot noise provides an alternative, purely electrical method of detection of these charge-neutral edge modes.
\end{abstract}
\maketitle

\section{Introduction}

\begin{figure*}[tb]
  \begin{center}
	 \includegraphics[width=0.9\linewidth]{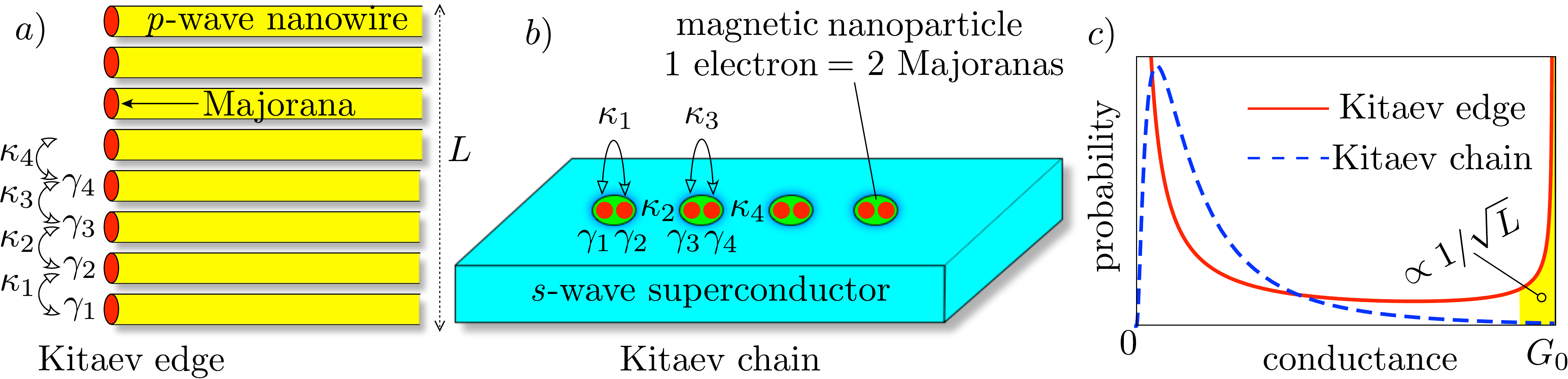}
  \end{center}
  \caption{Two realizations of the Kitaev Hamiltonian, at the edge of an array of  nanowires (Kitaev edge, panel \textit{a}) and in a chain of magnetic nanoparticles (Kitaev chain, panel \textit{b}, adapted from Refs.\ \onlinecite{Cho11,Nad13}). Statistical translational invariance at the Kitaev edge means that all couplings $\kappa_n$ between Majorana fermion operators $\gamma_{n}$ and $\gamma_{n+1}$ have the same statistical distribution. In the Kitaev chain it means that the couplings $\kappa_{2n}$ between nanoparticles have the same distribution, as well as the couplings $\kappa_{2n-1}$ of pairs of Majorana fermions within a nanoparticle --- while the sets $\kappa_{2n}$ and $\kappa_{2n-1}$ are unrelated. This difference is the reason that statistical translational invariance protects the Kitaev edge from localization, but not the Kitaev chain. As a consequence, the thermal conductance has a lognormal distribution in the Kitaev chain (dashed curve in panel \textit{c}), but a bimodal distribution in the 
Kitaev edge (solid curve, with a second peak of weight $\propto 1/\sqrt{L}$ at the conductance quantum $G_0$).
 }
  \label{fig_chainedge}
\end{figure*}

Gapless edge states are a striking manifestation of topological protection in two-dimensional systems. First studied in connection with the quantum Hall effect in a strong magnetic field \cite{Hal82,But88}, they are now known to exist also in the presence of time-reversal symmetry (for topological insulators) or particle-hole symmetry (for topological superconductors) \cite{Has10,Qi11}. The edge current can carry charge or heat, it can be uni-directional (chiral) or bi-directional (helical), but in each manifestation there is no backscattering --- so that the corresponding electrical or thermal conductance is quantized, independent of system size.

Isotropic two-dimensional superconductors with spin-triplet \textit{p}-wave pairing belong to the class of topological superconductors, with charge-neutral gapless edge states. Depending on the absence or presence of time-reversal symmetry, the edge modes can be chiral (class D) or helical (class DIII). It is known that the quantization of the thermal conductance breaks down if the two-dimensional superconductor is strongly anisotropic; the edge states remain, but backscattering by disorder is no longer forbidden by a topological invariant \cite{Asa12,Ful12,Ser14,Buh14}. One might surmise that the edge states will localize on length scales $L$ larger than the mean free path $\ell$, with an exponentially decaying conductance $\propto\exp(-L/\ell)$, but that is not what happens. Instead, in Ref.\ \onlinecite{Ful12} an anomalously slow (super-Ohmic) scaling $\propto \sqrt{\ell/L}$ was found, unlike that of any known one-dimensional system. A statistical symmetry (translational invariance of the disorder  
distribution) was identified as the origin of the topological protection \cite{Ful12}.

Here we study this remarkable delocalization of edge states in the framework of the Kitaev Hamiltonian \cite{Kit01} of randomly coupled Majorana fermions. We contrast the two realizations of the model illustrated in Fig.\ \ref{fig_chainedge}: at the edge of a two-dimensional superconductor (Kitaev edge, Fig.\ \ref{fig_chainedge}\textit{a}) and as a one-dimensional chain of nanoparticles (Kitaev chain, Fig.\ \ref{fig_chainedge}\textit{b}). While the Kitaev chain allows for delocalization, this requires a fine-tuning to the critical point of the topological phase diagram \cite{Bro00,Akh11}. Generically, the conductance of the Kitaev chain has a log-normal distribution peaked at an exponentially small value \cite{Mot01,Bro03,Gru05}, because disorder drives the system away from the gapless critical point into the gapped phase. In contrast, for the Kitaev edge we find a bimodal conductance distribution, with a second peak near the quantized conductance of the clean system. (Compare dashed and solid curves in Fig.\
 \ref{fig_chainedge}\textit{c}.) The $\sqrt{\ell/L}$ weight of this second peak produces the super-Ohmic conductance scaling of Ref.\ \onlinecite{Ful12}. 

We explain the difference in conductance distributions in terms of the different way in which translational invariance of the disorder distribution is realized in the two systems: in the Kitaev edge all nearest-neigbor coupling strengths of Majorana fermions are statistically equivalent, while in the Kitaev chain even and odd-numbered couplings are inequivalent. Finally, we show how the charge-neutral edge modes of the topological superconductor can be detected in an electrical --- rather than thermal --- measurement, by considering the shot noise of time-dependent current fluctuations.

The outline of this paper is as follows. In Section \ref{sec:models} we introduce model Hamiltonians for \textit{p}-wave superconductors with chiral or helical edge states and calculate the topological phase diagram in the presence of both anisotropy and disorder. The topological phase transitions are identified by considering the bulk conductance and the associated topological invariants. Edge conductance in the topologically nontrivial phases is studied in Section \ref{sec:edgeG}. In Section \ref{sec:kitaev} we contrast the conductance distributions of the Kitaev edge and the Kitaev chain. Electrical, rather than thermal, detection of the edge modes is discussed in Section \ref{electrdetect}. We conclude in Section \ref{conclude}.

\section{Topological phase diagrams of chiral and helical \textit{p}-wave superconductors}
\label{sec:models}

The topological phase diagram of clean chiral \textit{p}-wave superconductors (or superfluids) was studied in Refs.\ \onlinecite{Asa12,Ser14,Buh14}. Here we show how the topologically distinct phases evolve when we include disorder, for both chiral and helical pair potentials, which as we will see have a qualitatively different phase diagram. Numerical calculations on a disordered tight-binding model are compared with analytical calculations of the phase boundaries in self-consistent Born approximation.

\subsection{Model Hamiltonians}
\label{TBmodels}

Superconductors with broken spin-rotation symmetry are in symmetry class D or DIII in the Altland-Zirnbauer classification \cite{Alt97}, depending on whether time-reversal symmetry is broken or not. In both symmetry classes the Bogoliubov-De Gennes Hamiltonian ${\cal H}(\bm{k})$ has electron-hole symmetry,
\begin{equation}
\tau_x {\cal H}^{\vphantom{\dag}}(\bm k) \tau_x = -{\cal H}^{*}(-\bm k),\label{ehsymmetry}
\end{equation}
where the Pauli matrix $\tau_i$ acts on the electron-hole degree of freedom. In class DIII there is additionally the time-reversal symmetry
\begin{equation}
\sigma_y{\cal H}(\bm k)\sigma_y = {\cal H}^*(-\bm k),\label{Tsymmetry}
\end{equation}
with $\sigma_i$ acting on the spin degree of freedom and $\hbar\bm{k}$ the momentum.

The minimal class-D Hamiltonian, constrained by Eq.\ \eqref{ehsymmetry}, has the form
\begin{subequations}
\label{eq:HclassDk}
\begin{align}
&{\cal H}_{\rm D}(\bm k) = \epsilon(\bm k)\tau_z + \Delta_x\tau_x\sin k_x + \Delta_y\tau_y\sin k_y,\\
&\epsilon(\bm k) = -2t_x\cos k_x -2t_y\cos k_y -\mu,
\end{align}
\end{subequations}
where $(\Delta_{x},\Delta_{y})=(\Delta,\alpha\Delta)$ is the anisotropic amplitude of the chiral \textit{p}-wave pair potential (in a gauge where it's real), $(t_x,t_y)=(t,\alpha t)$ is the anisotropic hopping amplitude, and $\mu$ is the chemical potential. The parameter $\alpha\in[0,1]$ measures the degree of anisotropy, with $\alpha\rightarrow1$ the isotropic limit. We consider equal-spin pairing, so the spin degree of freedom does not appear in ${\cal H}_{\rm D}$.

In class DIII the additional constraint \eqref{Tsymmetry} is satisfied by taking two time-reversed copies of the Hamiltonian \eqref{eq:HclassDk},
\begin{align}
  {\cal H}_{\rm DIII}(\bm k) = {}& \epsilon(\bm k)(\sigma_0\otimes\tau_z) + \Delta_x(\sigma_z\otimes\tau_x)\sin k_x  \nonumber\\
  &+ \Delta_y(\sigma_0\otimes\tau_y )\sin k_y+ K(\sigma_y\otimes\tau_y),
  \label{eq:HclassDIIIk}
\end{align}
coupled with strength $K$.

\begin{figure}[tb]
  \begin{center}
	 \includegraphics[width=0.9\columnwidth]{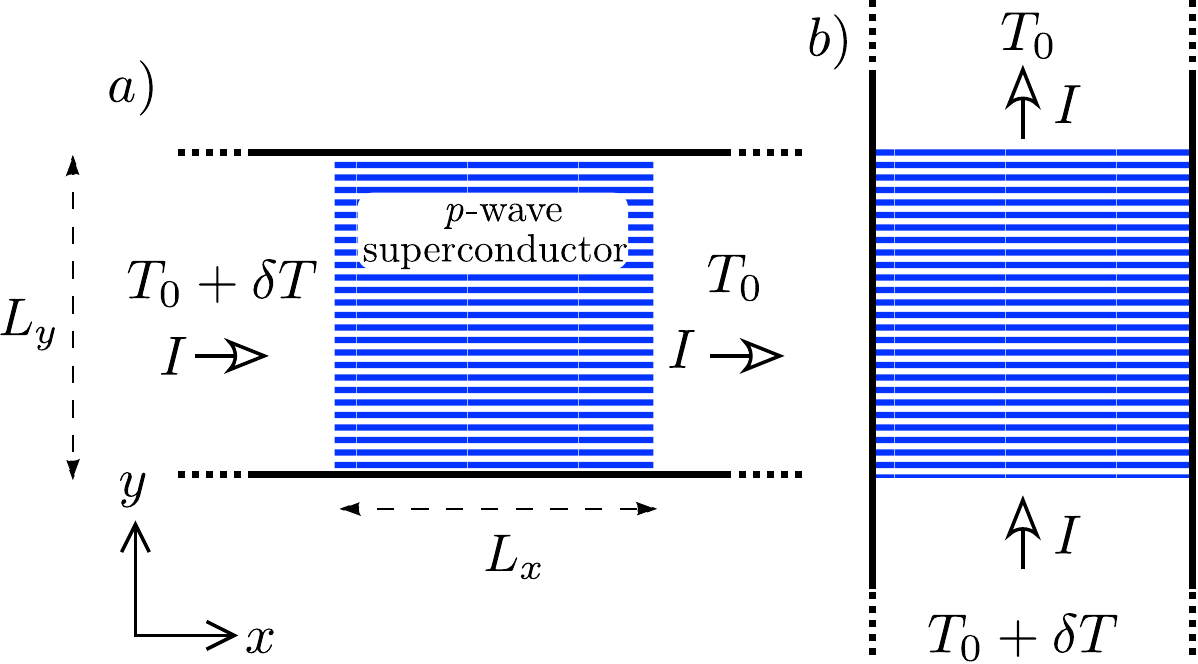}
  \end{center}
  \caption{Two-terminal geometries for thermal conduction in an anisotropic \textit{p}-wave superconductor. Panels \textit{a} and \textit{b} show the  perpendicular orientations of the heat current $I$ in response to a temperature difference $\delta T$. The thermal conductance in linear response is $G=\lim_{\delta T\rightarrow 0}I/\delta T$.
 }
  \label{fig_layout}
\end{figure}

The Hamiltonians \eqref{eq:HclassDk} and \eqref{eq:HclassDIIIk} are discretized on a two-dimensional square lattice of size $L_x\times L_y$ (lattice constant $a\equiv 1$). Electrostatic disorder (strength $\delta$) is added by randomly varying $\mu$, independently for each lattice site and uniformly in the interval $[\mu-\delta,\mu+\delta]$. We study thermal conduction by attaching disorder-free leads at two ends of the lattice, connected to reservoirs at temperature $T_0$ and $T_0+\delta T$ (see Fig.\ \ref{fig_layout}). The scattering matrix,\begin{equation}\label{eq:smatrix}
 S=\begin{pmatrix}
    r & t \\
    t' & r'
   \end{pmatrix},
\end{equation}
evaluated at the Fermi level ($E=0$) determines the thermal conductance
\begin{equation}
 G=G_0\,{\rm Tr}\,t^{\dagger}t,\;\;G_0=\pi^2k_{\rm B}^2T_0/6h,\label{Gthermaldef}
 \end{equation}
in the low-temperature, linear response regime. The numerical calculations are performed using the {\sc kwant} tight-binding code \cite{kwant}.

\subsection{Class D phase diagram}
\label{sec:modelD}

\subsubsection{Clean limit}

\begin{figure}[tb]
  \begin{center}
	 \includegraphics[width=0.8\columnwidth]{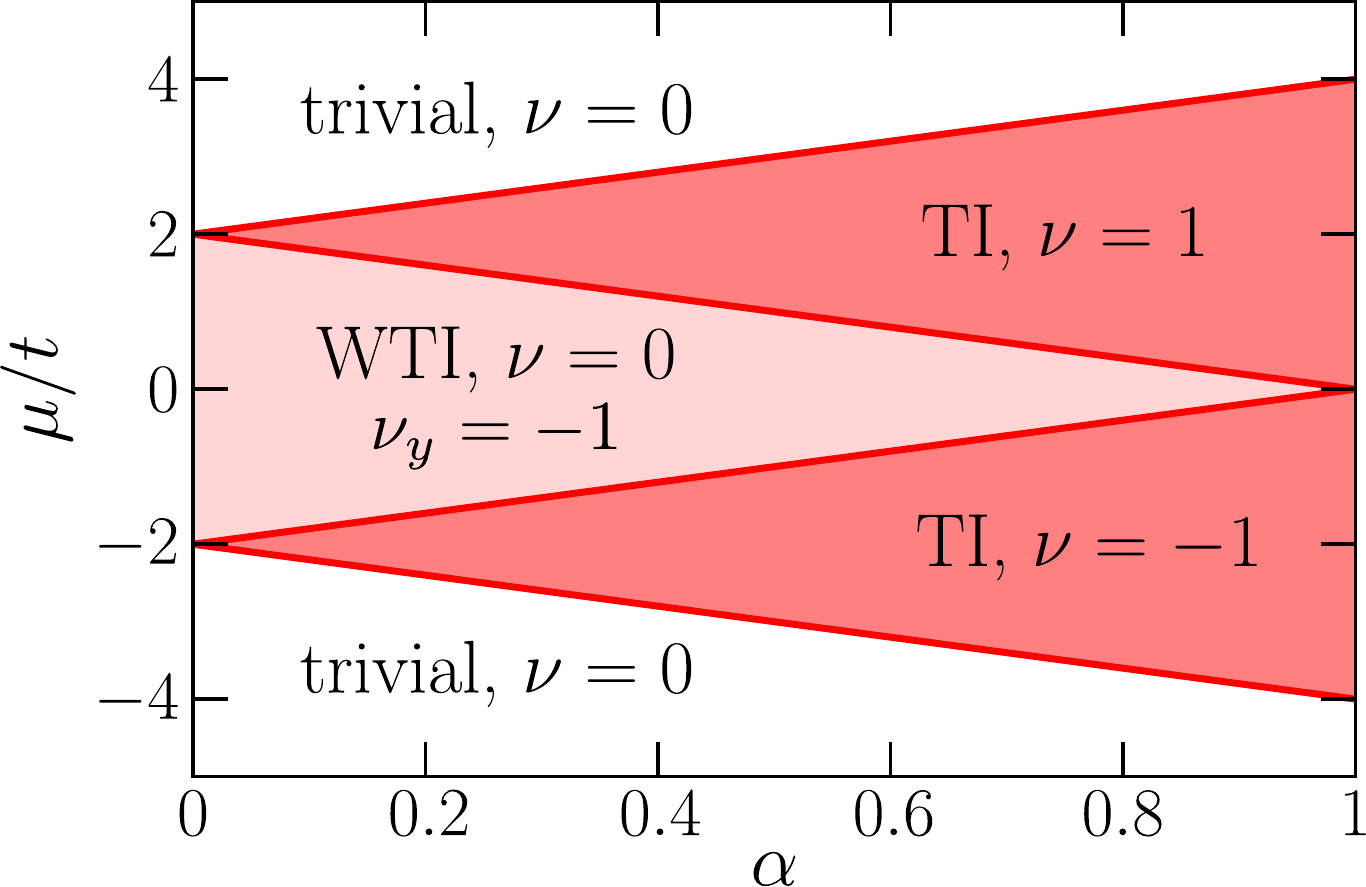}
  \end{center}
  \caption{Phase diagram of the class-D Hamiltonian~\eqref{eq:HclassDk} in the absence of disorder ($\delta=0$), as a function of chemical potential $\mu$ and anisotropy $\alpha$. The strong topological insulator phases (TI) have chiral Majorana modes along all edges, while the weak topological insulator phase (WTI) has Majorana modes only along edges oriented in the $y$-direction. The trivial phase has no edge modes.
  }
\label{fig:classDcleanphases}
\end{figure}

We first discuss the phase diagram of the class-D Hamiltonian \eqref{eq:HclassDk} in the clean limit of Refs.\ \onlinecite{Asa12,Ser14,Buh14}, before including the effects of disorder. Without disorder the momentum is a good quantum number and one can search for gap closings in the Brillouin zone. These occur at the four high-symmetry points $k_x,k_y\in\{0,\pi\}$, for chemical potentials $\mu=\pm 2t(1\pm\alpha)$. In the $\mu$-$\alpha$ plane the four gapless lines are boundaries separating five topologically distinct insulating phases, see Fig.\ \ref{fig:classDcleanphases}.

The number of chiral edge modes is given by the Chern number $\nu$ \cite{Has10,Qi11,Tho82}, being the winding number of eigenstates in the Brillouin zone. We compute this two-dimensional topological invariant from the scattering matrix rather than from an integral over the Brillouin zone, in a formulation that can be applied directly to disordered systems \cite{Bra10, Ful12c},
\begin{equation}\label{eq:d_invariant}
 \nu = \frac{1}{2\pi i}\int_0^{2\pi} d\phi \,\frac{d}{d\phi}\,\ln\,{\rm det}\,r(\phi).
\end{equation}
Here $r(\phi)$ is the reflection block of the scattering matrix \eqref{eq:smatrix} in the geometry of Fig.\ \ref{fig_layout}\textit{a}, with leads attached to $x=0$ and $x=L_x$ and twisted periodic boundary conditions \cite{note1} on the scattering state $\psi(x,y)$ in the $y$-direction: $\psi(x,0)=e^{i\phi}\psi(x,L_y)$.

For $2t(1-\alpha) < |\mu| < 2t(1+\alpha)$ the system is topologically non-trivial, with $\nu={\rm sign}\,\mu$ and a chiral Majorana edge mode. (The sign of $\nu$ gives the direction of propagation.) The absence of backscattering leads to a quantized thermal edge conductance $G=G_0$. This characterizes the strong topological insulator (TI). 

When $|\mu|>2t(1+\alpha)$ or $|\mu| < 2t(1-\alpha)$ the Chern number $\nu=0$, so there are no chiral edge modes. These regions in the phase diagram are distinguished by an alternative ``weak'' topological invariant $\nu_y$ \cite{Fu07,Fu07b}. Again, to prepare ourselves for disorder effects, we use a scattering matrix formulation rather than a Brillouin zone formulation \cite{Asa12,Ser14}. The two-dimensional weak topological invariant is the strong topological invariant in one dimension lower, which in class D is given by the determinant of the reflection matrix \cite{Akh11},
\begin{equation}\label{eq:d_invariant_weak}
 \nu_y = {\rm det}\,r(\phi=0).
\end{equation}
The dimensional reduction is implemented by evaluating $r(\phi)$ at $\phi=0$, so for periodic boundary conditions in the $y$-direction. When the Chern number $\nu=0$ the weak invariant $\nu_y$ may be equal to $+1$ (trivial insulator) or $-1$ (weak topological insulator, WTI).

For $|\mu|>2t(1+\alpha)$ we are in the topologically trivial phase, with $\nu_y=1$ and no edge modes at all. In contrast, when $|\mu| < 2t(1-\alpha)$ the system is a WTI, with $\nu_y=-1$ and non-chiral Majorana modes on the edges in the $y$-direction.

\subsubsection{Disorder effects}

\begin{figure}[tb]
  \begin{center}
	 \includegraphics[width=1\columnwidth]{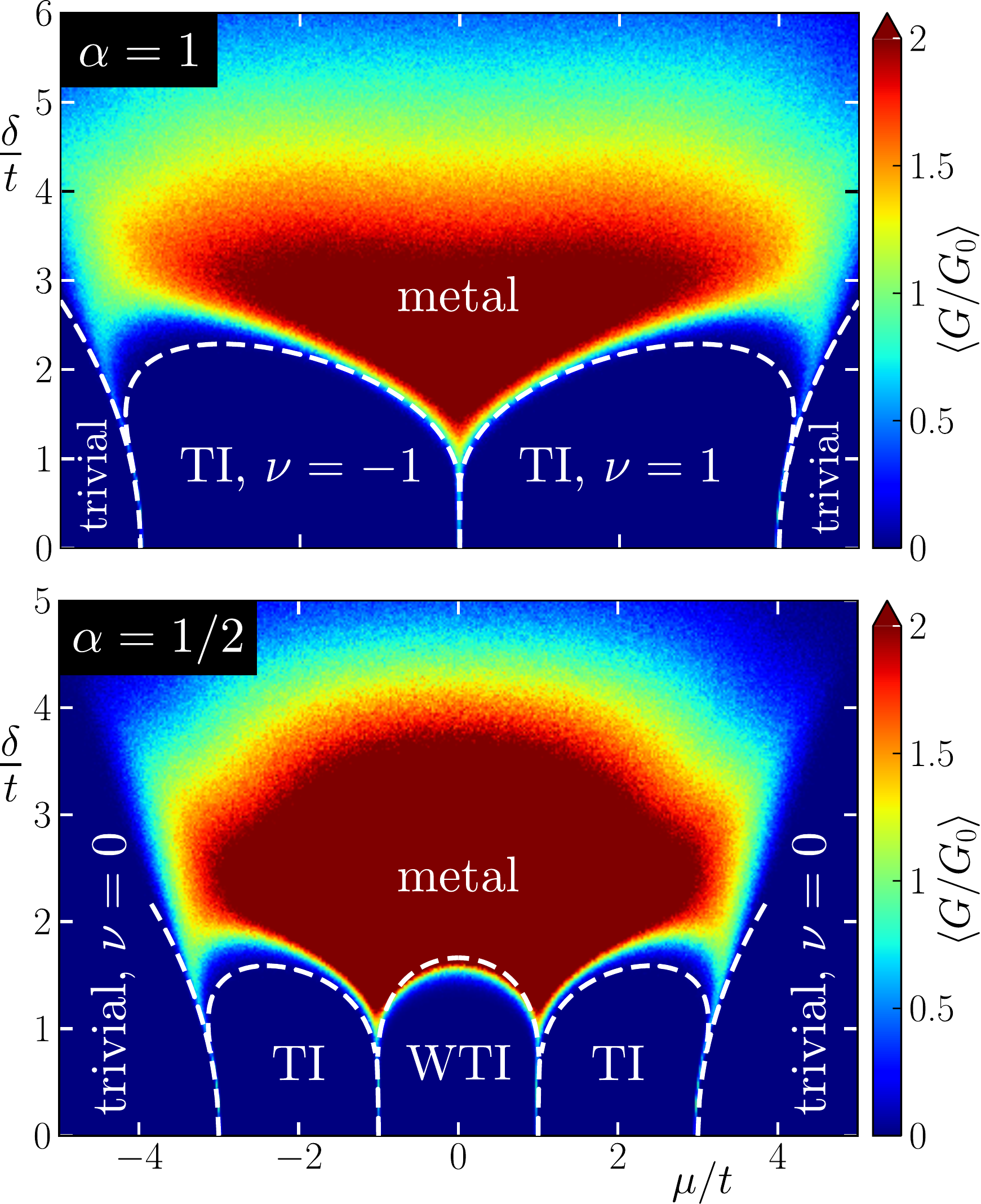}
  \end{center}
  \caption{Bulk thermal conductance for the class-D Hamiltonian \eqref{eq:HclassDk}, as a function of chemical potential $\mu$ and disorder strength $\delta$, for two values of the anisotropy at fixed $\Delta=t/2$. The data is averaged over 50 disorder realizations on a lattice of dimensions $L_x=L_y=50$ (current in the $x$-direction, periodic boundary conditions in the $y$-direction). The isotropic case (top panel, $\alpha=1$) shows gapped TI phases that are robust to disorder up to values $\delta\lesssim2t$. In the presence of anisotropy (bottom panel, $\alpha=1/2$), the weak topological insulator which forms at $|\mu|<2t(1-\alpha)$ survives up to disorder strengths of the same order as the TI phases. Dashed lines represent the phase boundaries in self-consistent Born approximation, without any fit parameter.
    }
  \label{fig:phases_d_disorder}
\end{figure}

Having described the phase diagram of the system in the clean limit, we now turn to the effects of disorder. Sufficiently strong disorder can convert a class-D superconductor that is insulating in the bulk into a thermal metal \cite{Sen00,Eve08,Med10}. To search for this topological phase transition we take the two-terminal geometry of Fig.\ \ref{fig_layout}\textit{a} with periodic boundary conditions in the $y$-direction, in order to focus on the metallic or insulating nature of the bulk. (We will consider edge conduction in Sec.\ \ref{sec:edgeG}.) 

Numerical results for the disorder-averaged thermal conductance $\langle G\rangle$ are shown in Fig.\ \ref{fig:phases_d_disorder}, as a function of chemical potential $\mu$ and disorder strength $\delta$. One can see that both the TI and WTI phases are robust to disorder, up to about $\delta \approx t$. For stronger disorder there is a TI-to-thermal metal phase transition, followed by a transition to a topologically trivial Anderson insulator. The phase boundaries between TI, WTI, and thermal metal are in quite good agreement with those calculated in self-consistent Born approximation (dashed lines in  Fig.\ \ref{fig:phases_d_disorder}, see App.\ \ref{scBa} for details of the calculation). The transition to an Anderson insulator at strong disorder is out of reach of that approximation. 

\begin{figure}[tb]
  \begin{center}
	 \includegraphics[width=0.9\columnwidth]{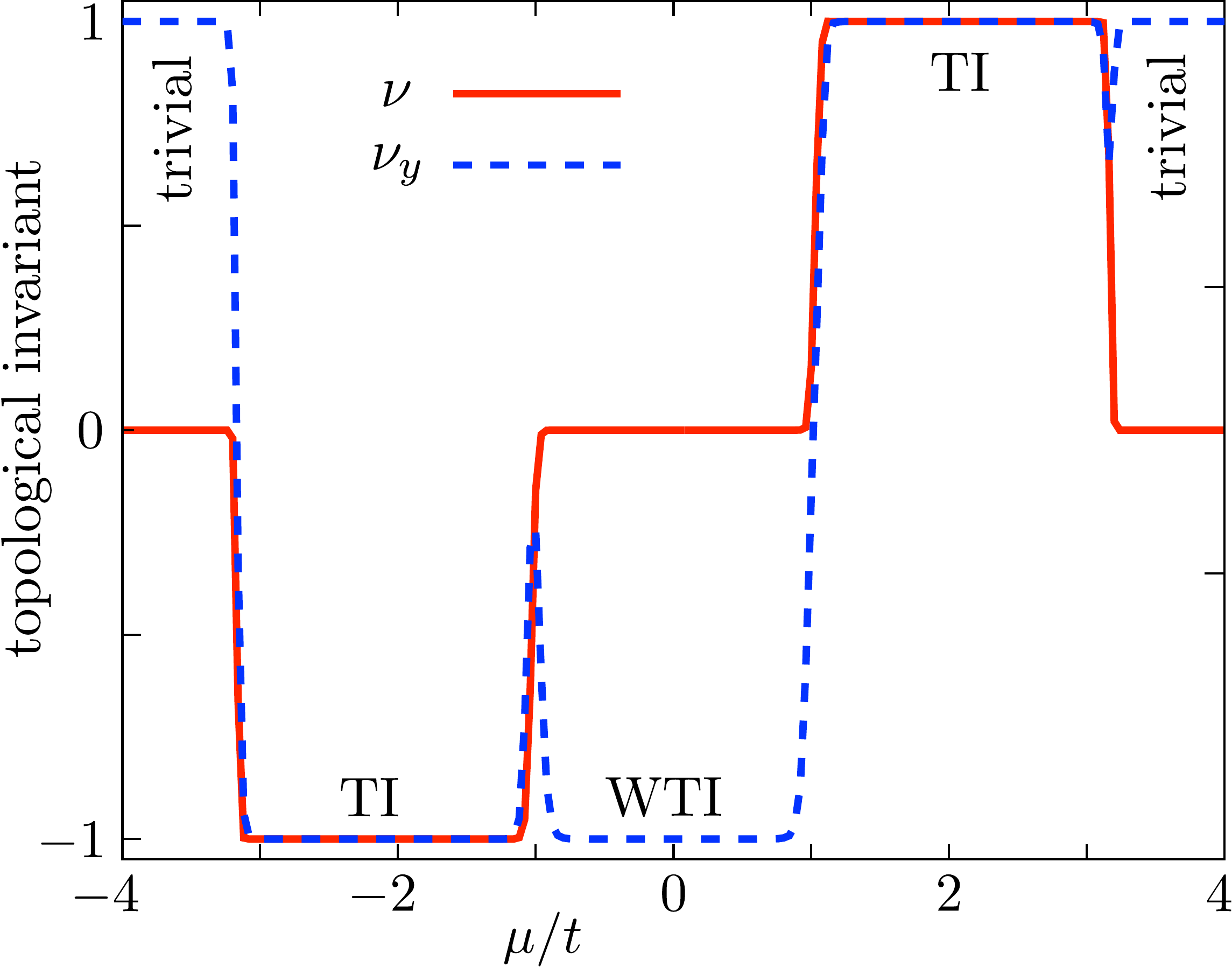}
  \end{center}
  \caption{Topological phase transitions signaled by a change in the Chern number $\nu$ (red solid line) or the weak invariant $\nu_y$ (blue dashed line). The curves are calculated from the class-D Hamiltonian \eqref{eq:HclassDk}, using the scattering matrix formulas \eqref{eq:d_invariant} and \eqref{eq:d_invariant_weak}, for $\Delta=t/2$, $\alpha=1/2$, $\delta=t$, averaged over 4000 disorder realizations in a system of size $L_x\times L_y = 50\times50$.}
  \label{fig:dQ}
\end{figure}

The distinct topological nature of the TI, WTI, and trivial phase is confirmed by a calculation of the topological invariants $\nu,\nu_y$, see Fig.\ \ref{fig:dQ}. In the bulk insulating phases these are quantized numbers: $\nu\in\{-1,0,1\}$ (a so-called $\mathbb{Z}$ invariant), while $\nu_y\in\{-1,1\}$ (a $\mathbb{Z}_2$ invariant). At the topological phase transitions, when the bulk gap closes, both $\nu$ and $\nu_y$ are free to vary between these integer values, resulting in the smooth transitions shown in Fig.\ \ref{fig:dQ}.

\subsection{Class DIII phase diagram}
\label{sec:modelDIII}

\subsubsection{Clean limit}

We now turn to the phase diagram of the class DIII Hamiltonian \eqref{eq:HclassDIIIk}, first without disorder. It is convenient to rotate the Hamiltonian to a block off-diagonal form,
\begin{subequations}
\label{eq:offdiagHDIII}
\begin{align}
& U {\cal H}_{\rm DIII} U^\dag = \begin{pmatrix}
                          0 & A \\
			  A^\dag & 0
                         \end{pmatrix},\;\;U=\exp(-\tfrac{1}{4}i\pi \,\sigma_x\otimes\tau_x),\\
                          &A =  i\Delta\sigma_0\sin k_x+\alpha\Delta\sigma_z\sin k_y\nonumber\\
                          &\quad\quad\mbox{}+ (\mu+2t\cos k_x+2\alpha t\cos k_y-iK)\sigma_y.\label{eq:DHIIIblock}
\end{align}
\end{subequations}
At the gap closings of $H_{\rm DIII}$ the determinant of $A$ vanishes, which happens when
\begin{subequations}
\label{eq:diii_gap_closing}
\begin{align}
&\mu/t = -2\cos k_x - 2\alpha \cos k_y,\label{eq:diii_gap_closing1}\\
& K^2/\Delta^2 = \sin^2 k_x + \alpha^2\sin^2 k_y.\label{eq:diii_gap_closing2}
\end{align}
\end{subequations}

\begin{figure}[tb]
  \begin{center}
	 \includegraphics[width=0.8\columnwidth]{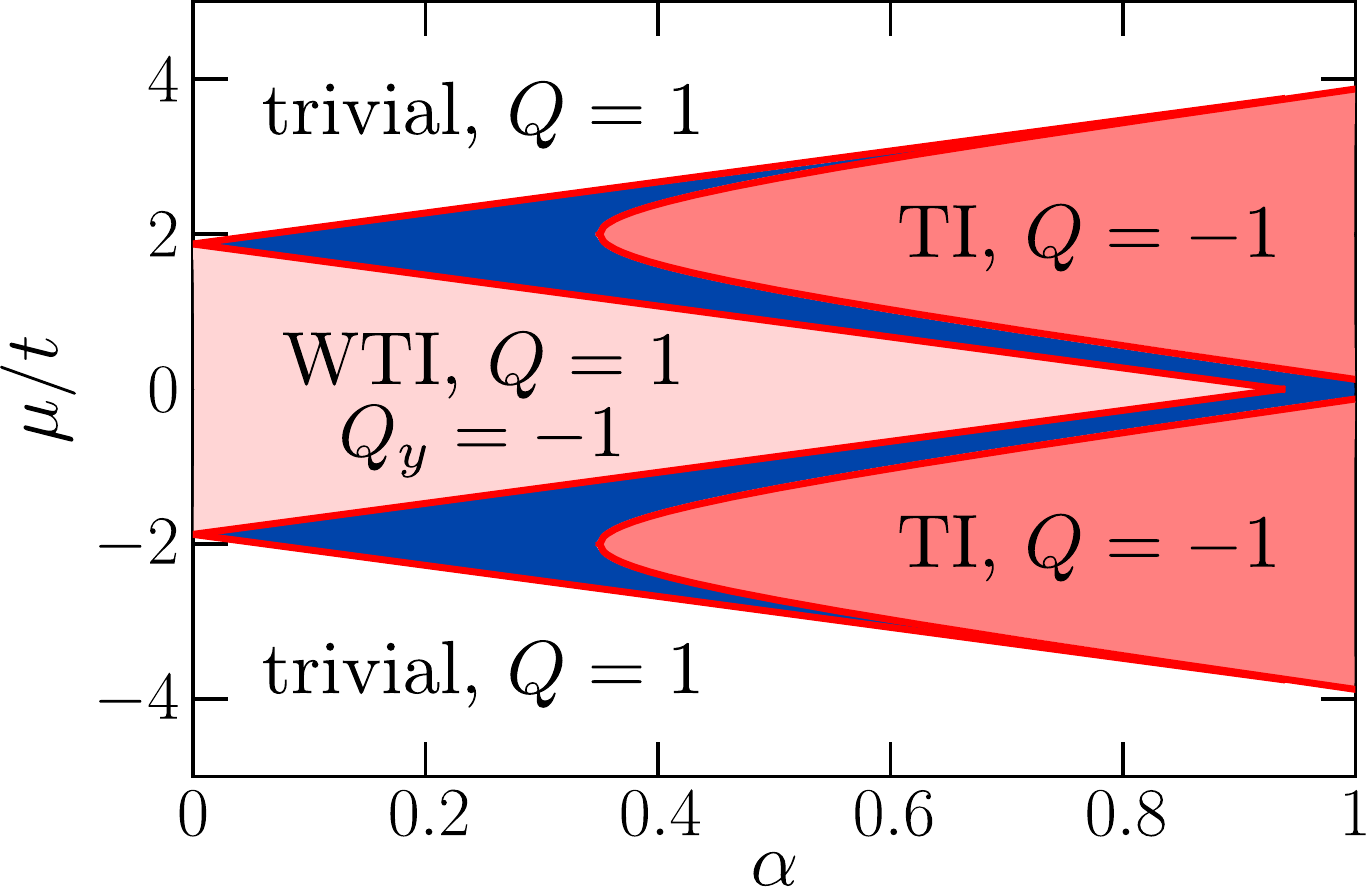}
  \end{center}
  \caption{Phase diagram of the class-DIII Hamiltonian \eqref{eq:HclassDIIIk} without disorder, for $K/\Delta=0.35$. Topologically distinct insulating phases are separated by gapless metallic regions (blue). The topologically trivial insulator, without edge states, exists for any amount of anisotropy, while the TI and WTI phases with edge states require, respectively, $\alpha > K/\Delta$ and $\alpha < \sqrt{1-K^2/\Delta^2}$.}
  \label{fig:classDIIIcleanphases}
\end{figure}

The gap closings identify the boundaries of insulating phases, as shown in Fig.\ \ref{fig:classDIIIcleanphases}. While in class D the gap closes along a line in the phase diagram, in class DIII there are extended gapless regions of a metallic phase separating the insulating phases. (This is a generic feature of helical \textit{p}-wave superconductors \cite{Ful12b}.)

We can distinguish five distinct insulating regions of phase space. For weak anisotropy, $\alpha > K/\Delta$, we find two $\pm\mu$ symmetric insulating phases bounded by
\begin{equation}\label{eq:mu_ti}
 2-2\sqrt{\alpha^2-K^2/\Delta^2} < |\mu/t| < 2+2\sqrt{\alpha^2-K^2/\Delta^2}.
\end{equation}
When the anisotropy reaches the critical value
\begin{equation}
\alpha_c=(1 - K^2 / \Delta^2)^{1/2}, 
\end{equation}a third insulating phase appears centered around $\mu=0$, in the interval
\begin{equation}\label{eq:mu_sti}
 |\mu/t| < 2(\alpha_c-\alpha),\;\;\text{for}\;\;\alpha<\alpha_c.
\end{equation}
Additionally, for any amount of anisotropy there are insulating phases at large chemical potentials, with boundaries given by
\begin{equation}\label{eq:mu_triv}
 |\mu/t| <\begin{cases}
   2\alpha+2\sqrt{1-K^2/\Delta^2}, &\text{for}\;\; \alpha<\alpha_c \\
   2\sqrt{2}\sqrt{1+\alpha^2-K^2/\Delta^2}, &\text{for}\;\; \alpha>\alpha_c.
\end{cases}
\end{equation}

To test the topological properties of these phases, we compute the associated DIII topological invariants in a scattering formulation (so that we can directly apply it to disordered systems in the next subsection). The strong topological invariant \cite{Ful12c},
\begin{equation}\label{eq:diii_invariant}
 Q={\rm Pf}\,[i\sigma_y r(\phi=0)]\times{\rm Pf}\,[i\sigma_y r(\phi=\pi)],
\end{equation}
is determined by the Pfaffians of the reflection matrix with periodic ($\phi=0$) and anti-periodic ($\phi=\pi$) boundary conditions in the $y$-direction. In view of the time-reversal symmetry condition \eqref{Tsymmetry}, the matrix $i\sigma_y r(\phi)$ is antisymmetric for $\phi=0,\pi$, so the Pfaffian exists.

The insulating regions delimited by Eq.~\eqref{eq:mu_ti} are topologically non-trivial ($Q=-1$), with helical Majorana edge states and quantized thermal conductance $G=2G_0$. All other phases have $Q=1$. The ones appearing at large chemical potentials, bounded by Eq.~\eqref{eq:mu_triv}, are topologically trivial, without edge states. However, the phase which develops at the critical anisotropy $\alpha_c$, bounded by Eq.~\eqref{eq:mu_sti}, has $Q=1$ but still supports gapless modes on edges oriented in the $y$-direction. The weak topological invariant $Q_y=-1$ of this phase is obtained by dimensional reduction to the one-dimensional class-DIII topological invariant \cite{Ful11},
\begin{equation}\label{eq:diii_invariant_weak}
 Q_y = {\rm Pf}\,[i\sigma_{y}r(\phi=0)].
\end{equation}

\subsubsection{Disorder effects}

\begin{figure}[tb]
  \begin{center}
	 \includegraphics[width=1\columnwidth]{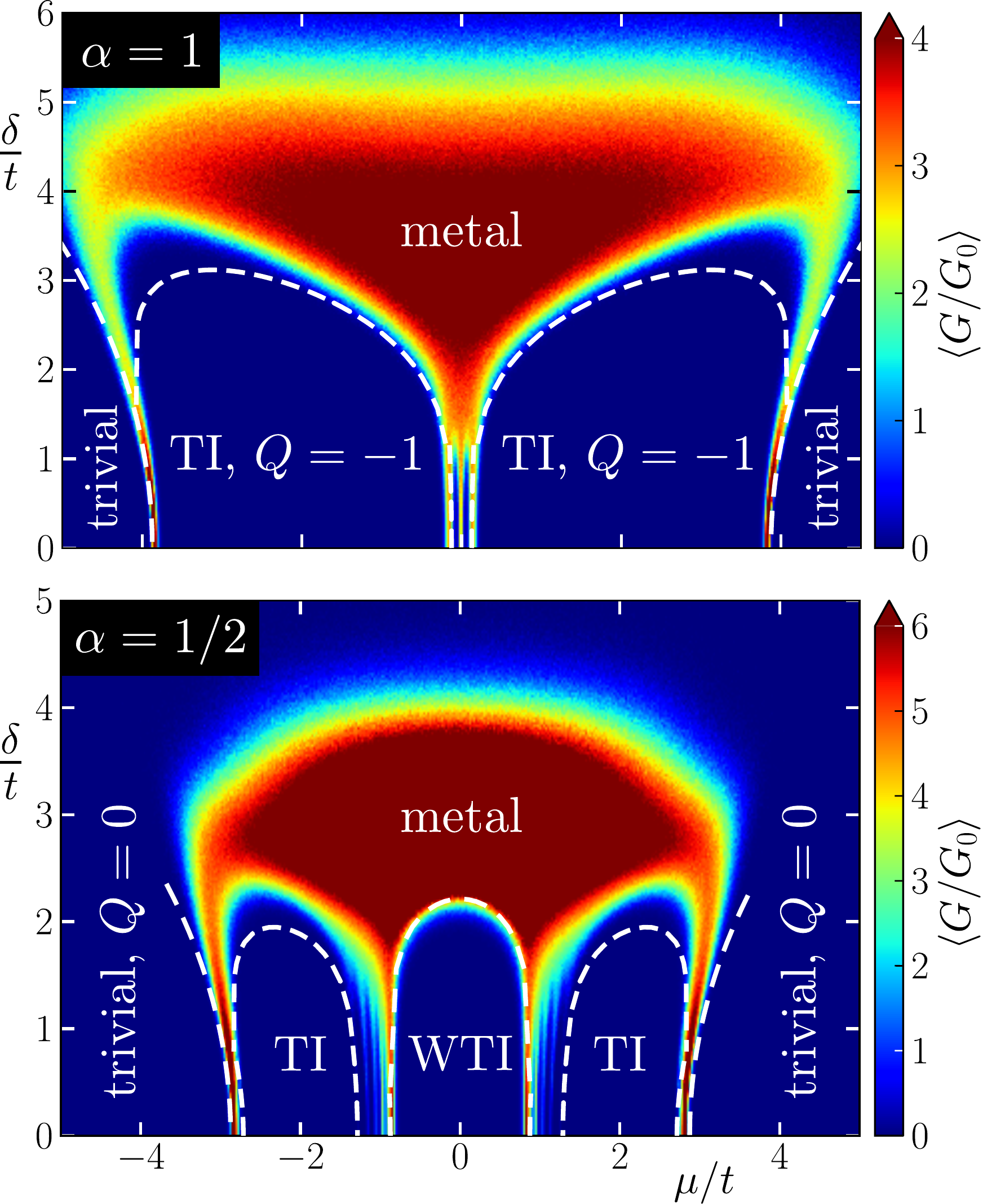}
  \end{center}
  \caption{Same as Fig.\ \ref{fig:phases_d_disorder} for the class-DIII Hamiltonian \eqref{eq:HclassDIIIk} (with $\Delta=t$, $K=0.35\,t$, other parameters unchanged). Notice that the thermal metal phase starts already at zero disorder.
  }
  \label{fig:diii_phases_disorder}
\end{figure}

Fig.\ \ref{fig:diii_phases_disorder} shows the effect of disorder on the topological phases, probed by calculating the thermal conductance in the geometry of Fig.\ \ref{fig_layout}\textit{a} with periodic boundary conditions in the $y$-direction. Comparison with the class-D phase diagram of Fig.\ \ref{fig:phases_d_disorder} shows as a qualitative difference that the thermal metal phase extends down to zero disorder. This behavior is also captured by the self-consistent Born approximation (dashed curves), see App.\ \ref{scBa} for details of the calculation.

\begin{figure}[tb]
  \begin{center}
	 \includegraphics[width=0.8\columnwidth]{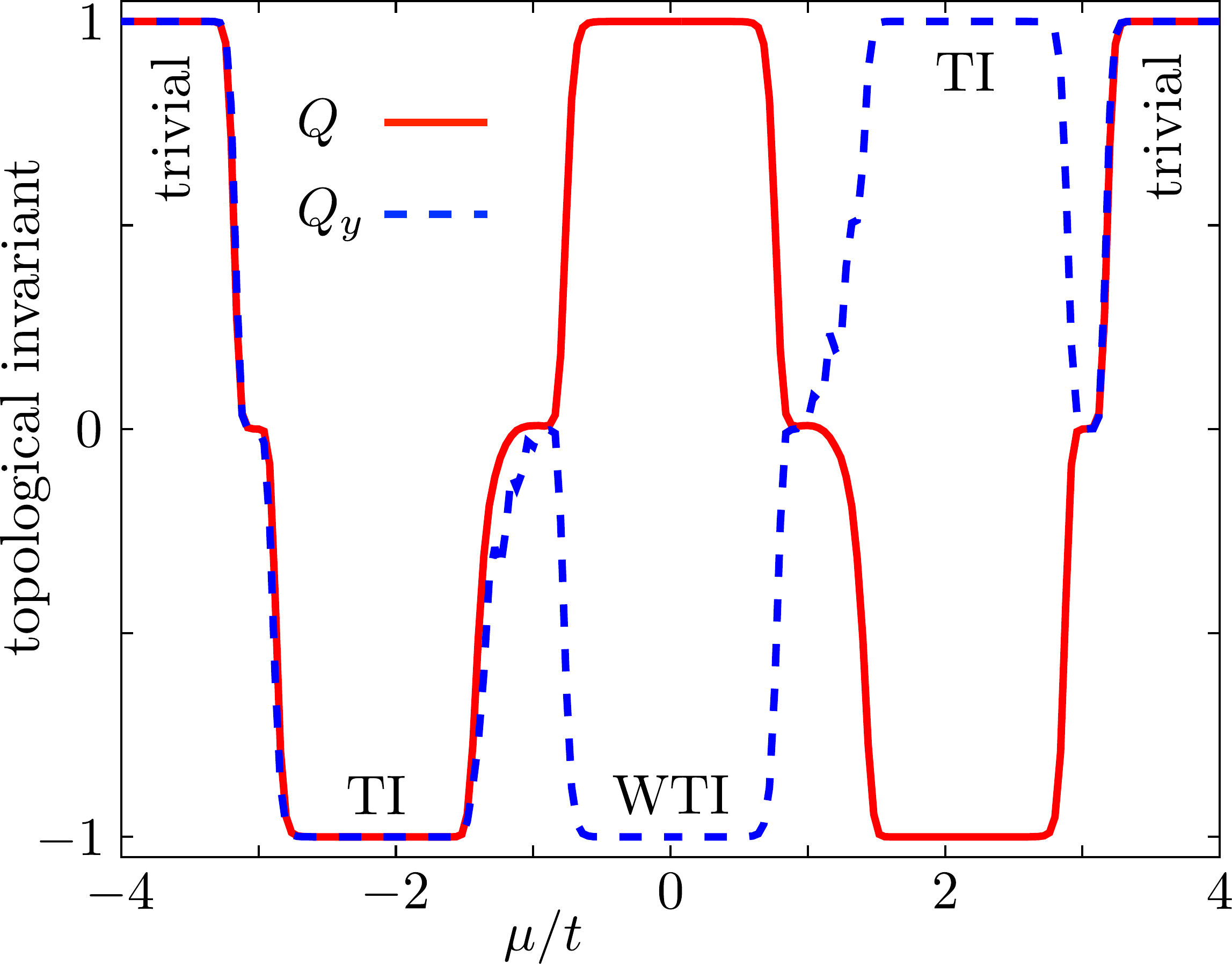}
  \end{center}
  \caption{Same as Fig.\ \ref{fig:dQ} for the class-DIII topological invariants $Q$, Eq.\ \eqref{eq:diii_invariant}, and $Q_y$, Eq.\ \eqref{eq:diii_invariant_weak} (with $\delta=1.5\,t$, other parameters as in Fig.\ \ref{fig:diii_phases_disorder}).
  }
  \label{fig:diiiQ}
\end{figure}

The scattering matrix formulas \eqref{eq:diii_invariant} and \eqref{eq:diii_invariant_weak} for the class-DIII strong and weak topological invariants are applied to a disordered system in Fig.\ \ref{fig:diiiQ}. These are both $\mathbb{Z}_2$ invariants, meaning that they take on the values $\pm 1$ when the bulk is insulating: $Q=Q_y=1$ in the trivial insulator, $Q=-1$, $Q_y=\pm 1$ in the TI, and $Q=1$, $Q_y=-1$ in the WTI.

\section{Edge conduction}
\label{sec:edgeG}

So far we studied thermal conduction in the geometry of Fig.\ \ref{fig_layout}\textit{a} with periodic boundary conditions in the transverse direction, in order to eliminate edge contributions and focus on bulk properties. To study edge conduction in the TI and WTI phases we now take the geometry of Fig.\ \ref{fig_layout}\textit{b}, with leads attached to $y=0,L_y$ and hard-wall boundary conditions at $x=0,L_x$. We again first consider the clean case and then add the effects of disorder, for both symmetry classes D and DIII.

\subsection{Clean case}

\begin{figure}[tb]
  \begin{center}
	 \includegraphics[width=\columnwidth]{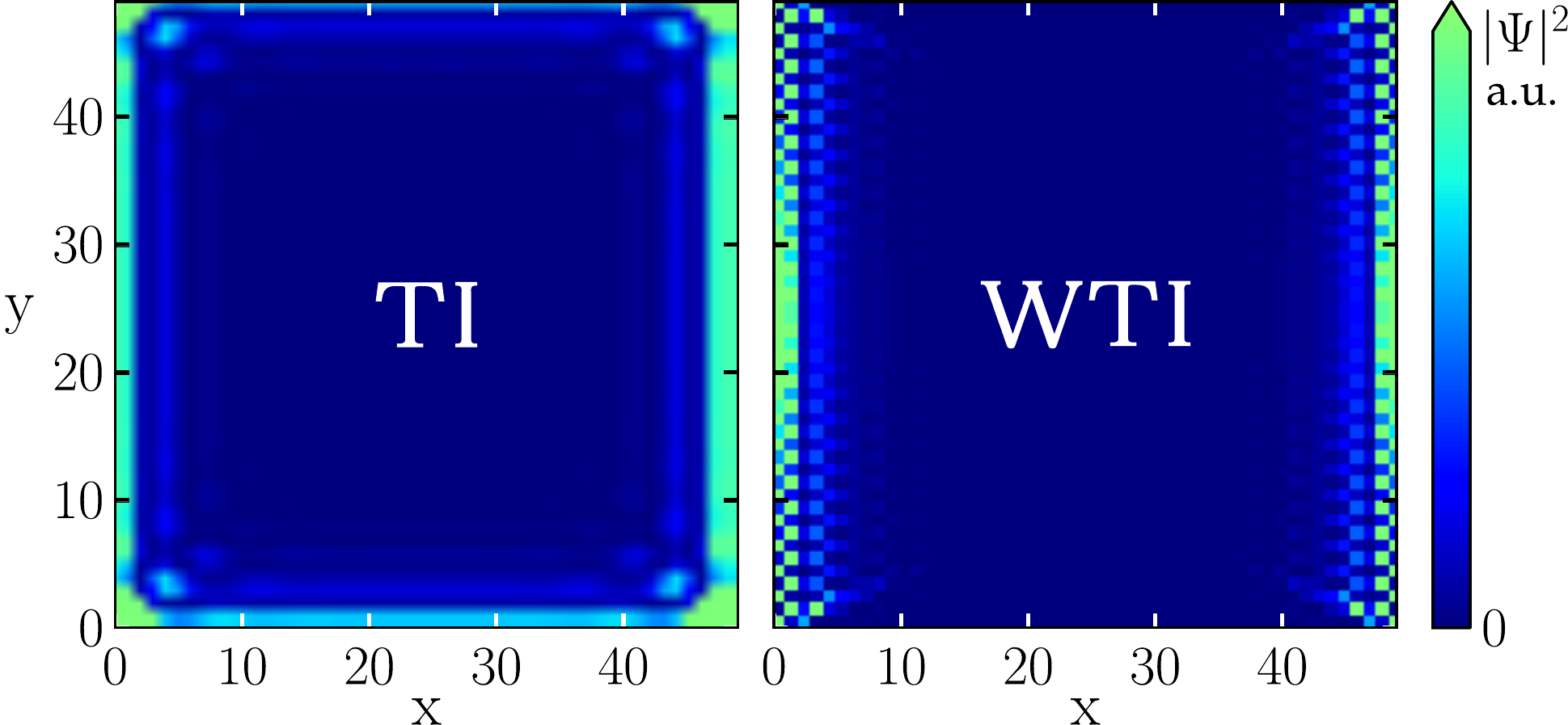}
  \end{center}
  \caption{Intensity of the lowest eigenstate of the class-D Hamiltonian \eqref{eq:HclassDk} (with $\delta=0$, $\alpha=1/2$, $\Delta=t/2$). The edge states in the TI ($\mu=2.2\,t$) and WTI ($\mu=0.2\,t$) phases are contrasted in the two panels.}
  \label{fig:ldos}
\end{figure}

The TI and WTI phases both have gapless edge states, the difference being that the TI edge states appear on all edges while the WTI edge states exist only at two of the four edges (see Fig.\ \ref{fig:ldos}). In the geometry of Fig.\ \ref{fig_layout}\textit{b} we can probe the edge conductance in both phases. Without disorder the conductance is system-size independent, because there is no backscattering, and the difference between the TI and WTI phases is simply a factor of two. 

\begin{figure}[tb]
  \begin{center}
	 \includegraphics[width=0.8\columnwidth]{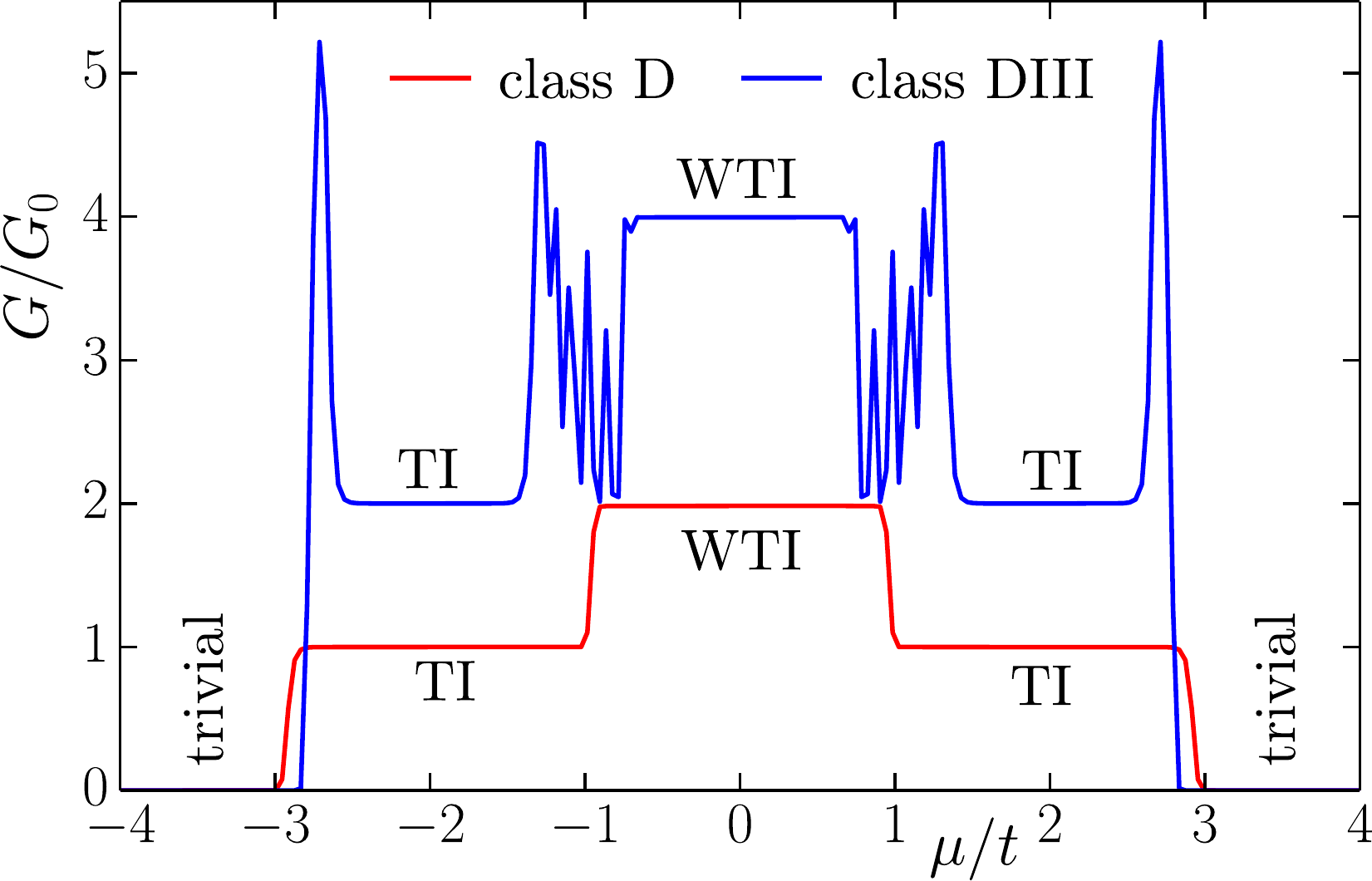}
  \end{center}
  \caption{Thermal conductance without disorder in class D and DIII. The geometry is that of Fig.~\ref{fig_layout}\textit{b} with dimensions $L_x=L_y=50$ and hard-wall boundaries in the $x$-direction. Parameters are $\Delta=t$, $\alpha=1/2$, and $K=0.35\,t$. The transition from the TI to the WTI phase is marked by a doubling of the edge conductance in the absence of backscattering. In class DIII the transition occurs via an intermediate region of thermal conduction through the gapless bulk.}
  \label{fig:cleanEdgeG}
\end{figure}

This conductance doubling at the TI-to-WTI transition is shown in Fig.\ \ref{fig:cleanEdgeG}. In class D it happens because the chiral edge state of the TI phase can propagate in both directions in the WTI phase. In class DIII we start out with helical edge states in the TI phase, with direction of propagation tied to the spin degree of freedom. In the WTI phase this helicity is lost, so now both spin-up and spin-down can propagate in both directions and the conductance is doubled.

\subsection{Disorder effects}

\begin{figure}[tb]
  \begin{center}
	 \includegraphics[width=0.8\columnwidth]{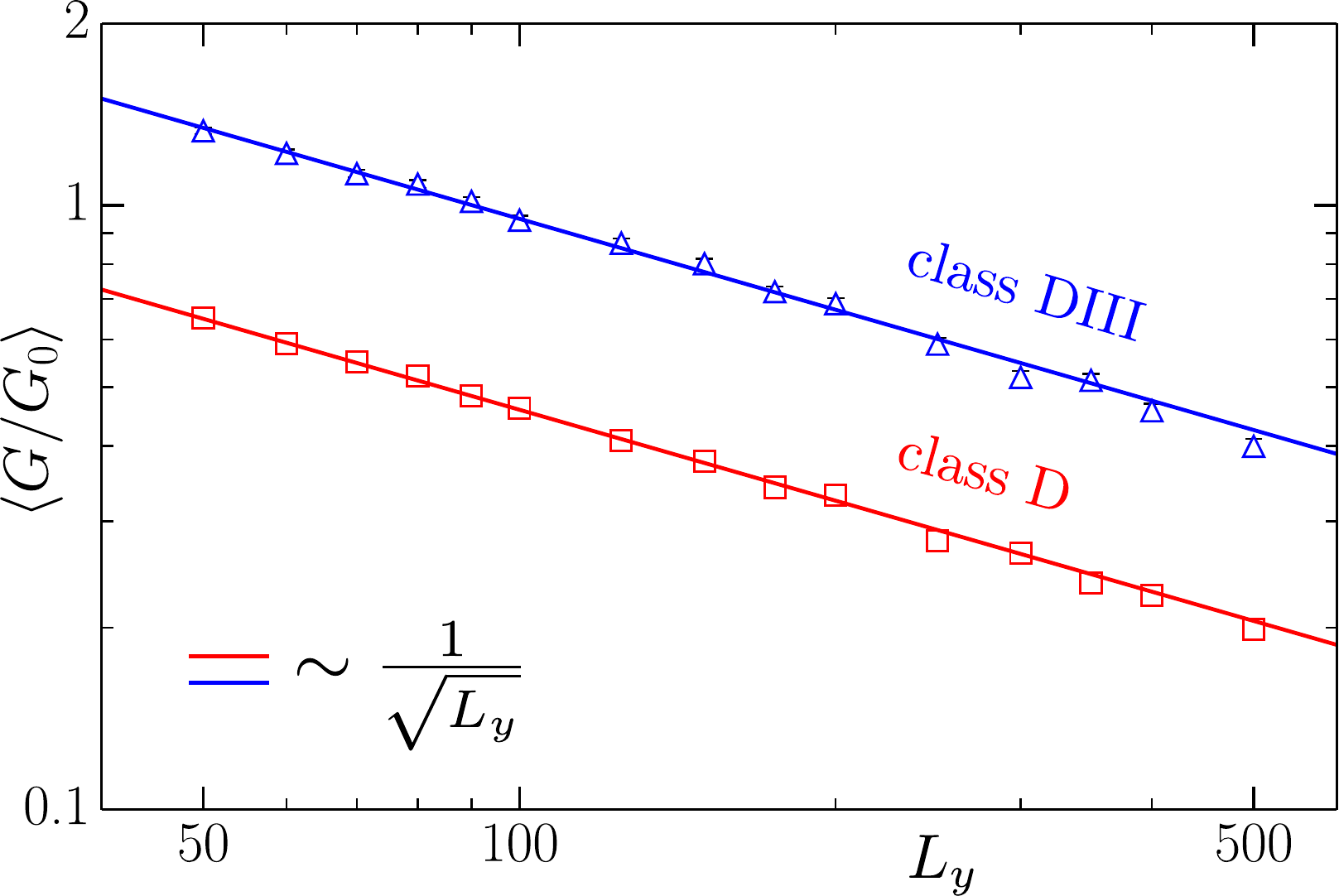}
  \end{center}
  \caption{Disorder-averaged thermal conductance in the WTI phase of class D and DIII, at fixed $L_x=50$ and varying $L_y$. Data points are averaged over 4000 disorder realizations, for $\mu=0$, $\delta=1.5\,t$, other parameters as in Fig.\ \ref{fig:cleanEdgeG}. Solid lines show the expected $L_y^{-1/2}$ scaling in the log-log plot.
  }
  \label{fig:DandDIIIscaling}
\end{figure}

The addition of disorder has no effect on the conductance in the TI phase, since backscattering of chiral or helical edge states is forbidden. The edge states in the WTI are neither chiral nor helical, so disorder does cause backscattering and reduces the edge conductance. However, as discovered in Ref.\ \onlinecite{Ful12}, the recovery of translational invariance upon ensemble averaging prevents localization of the WTI edge states. Instead of an exponential decay of the conductance with the length $L\equiv L_y$ of the edge, there is only an algebraic $1/\sqrt{L}$ decay. In Fig.\ \ref{fig:DandDIIIscaling} we show this super-Ohmic conductance scaling for the WTI phase of the class-D and class-DIII Hamiltonians (\ref{eq:HclassDk}) and (\ref{eq:HclassDIIIk}).

\section{Kitaev chain versus Kitaev edge}
\label{sec:kitaev}

The absence of localization at the edge of the anisotropic \textit{p}-wave superconductor is puzzling if one tries to understand it starting from the limit $\alpha\rightarrow 0$ of strong anisotropy. Then the system can be thought of as an array of weakly coupled nanowires with overlapping Majorana zero-modes at the end points, a so-called Kitaev ladder \cite{Wan14,Wak14,Dum14}. The effective edge Hamiltonian is the Kitaev Hamiltonian \cite{Kit01} in class D, or two time-reversed copies of it in class DIII. The disordered one-dimensional Kitaev model, called the Kitaev chain, is known to be an insulator \cite{Mot01,Bro03,Gru05} --- so how do the Kitaev edge modes avoid localization? 

To clarify the situation we contrast the two class-D systems. (Class DIII is similar.) The Kitaev Hamiltonian
\begin{equation}
H_{\rm K}=\sum_{n=1}^{2N}i\kappa_n\gamma_{n}\gamma_{n+1}\label{HKdef}
\end{equation}
describes the nearest-neigbor coupling (coupling strength $\kappa_n$) of $2N$ Majorana fermion operators $\gamma_{n}$. These are Hermitian operators, $\gamma_n=\gamma_n^{\dagger}$, with anti-commutation relation $\gamma_n\gamma_m+\gamma_m\gamma_n=2\delta_{nm}$. To obtain a closed system the Majorana's are assumed to lie on a ring, so that $\kappa_{2N}$ couples $\gamma_{2N}$ to $\gamma_{2N+1}\equiv\gamma_1$.

This one-dimensional system in symmetry class D has a $\mathbb{Z}_2$ topological invariant \cite{Kit01}, 
\begin{equation}
Q_{\rm K}={\rm sign}\,({\rm Pf}\,A^{+})({\rm Pf}\,A^{-}),\label{QKdef}
\end{equation}
determined by the Pfaffians of a pair of real antisymmetric matrices $A^{\pm}$, having nonzero matrix elements
\begin{subequations}
\label{Askewdef}
\begin{align}
&A_{n,n+1}^{\pm}=-A_{n+1,n}^{\pm}=\kappa_{n},\;\;1\leq n\leq 2N-1,\label{Askewdefa}\\
&A_{2N,1}^{\pm}=-A_{1,2N}^{\pm}=\pm\kappa_{2N}.\label{Askewdefb}
\end{align}
\end{subequations}
Evaluation of the Pfaffians gives
\begin{equation}
Q_{\rm K}={\rm sign}\,\left(\prod_{n=1}^{N}\kappa_{2n-1}^2-\prod_{n=1}^{N}\kappa_{2n}^2\right).\label{QKresult}
\end{equation}

Translational invariance of the disorder ensemble means two completely different things for the Kitaev edge and for the Kitaev chain. For the Kitaev edge it means that the coupling strengths $\kappa_n$ between adjacent Majorana's all have the same distribution. The disorder average $\langle Q_{\rm K}\rangle$ of the topological invariant then vanishes, which is why the Kitaev edge is called a \textit{critical} WTI \cite{Ful12}. In contrast, as illustrated in Fig.\ \ref{fig_chainedge}, for the Kitaev chain translational invariance means that the $\kappa_n$'s with $n$ even or those with $n$ odd have the same distribution, but the distributions of $\kappa_{2n}$ and $\kappa_{2n-1}$ are unrelated. The topological invariant is then nonzero on average, so the Kitaev chain is {\em not critical} \cite{Mot01,Gru05}.

The implication for the transmission probability $T$ follows if we remove the coupling between $\gamma_{2N}$ and $\gamma_1$, so that we can introduce transmission and reflection amplitudes $t=1/\cosh\alpha$, $r=\tanh\alpha$. The Lyapunov exponent $\alpha$ determines both $T$ and $Q_{\rm K}$,
\begin{equation}
Q_{\rm K}={\rm sign}\,\alpha,\;\;T=1/\cosh^2\alpha,\label{QKGrelation}
\end{equation}
and has a Gaussian distribution $P(\alpha)$ \cite{Bro00,Gru05}. The variance ${\rm Var}\,\alpha=L/\ell$ is determined by the mean free path $\ell$ for backscattering along the edge, of length $L\gg\ell$. The mean $\langle\alpha\rangle=L/\xi$ defines the localization length $\xi$. A vanishing $\langle Q_{\rm K}\rangle$ implies that the median of $P(\alpha)$ is zero, and since it's Gaussian also $\langle\alpha\rangle=0\Rightarrow\xi=\infty$.

For the Kitaev chain $\langle Q_{\rm K}\rangle\neq 0$ and hence $\xi$ is finite, so the transmission probability has a log-normal distribution peaked at $T=e^{-2L/\xi}$, with an exponentially decaying average transmission \cite{Mot01,Gru05}. In contrast, for the Kitaev edge $\langle Q_{\rm K}\rangle=0\Rightarrow \langle\alpha\rangle=0\Rightarrow\xi=\infty$. The Gaussian distribution of the Lyapunov exponent then produces a {\em bimodal} distribution of the transmission probability for $L\gg \ell$,
\begin{align}
P(T)&=\int_{-\infty}^{\infty}d\alpha\,\delta(T-1/\cosh^2\alpha)(2\pi L/\ell)^{-1/2}e^{-\alpha^2\ell/2L}\nonumber\\
&=(\ell/2\pi L)^{1/2}\,T^{-1}(1-T)^{-1/2}\nonumber\\
&\qquad\times\exp\bigl[-(\ell/2L)\,{\rm arcosh}^{2}(T^{-1/2})\bigr],\label{PTbimodal}
\end{align}
peaked near $T=0$ and $T=1$, with average decaying algebraically as $\langle T\rangle=\sqrt{2\ell/\pi L}$.

\begin{figure}[tb]
  \begin{center}
	 \includegraphics[width=0.9\columnwidth]{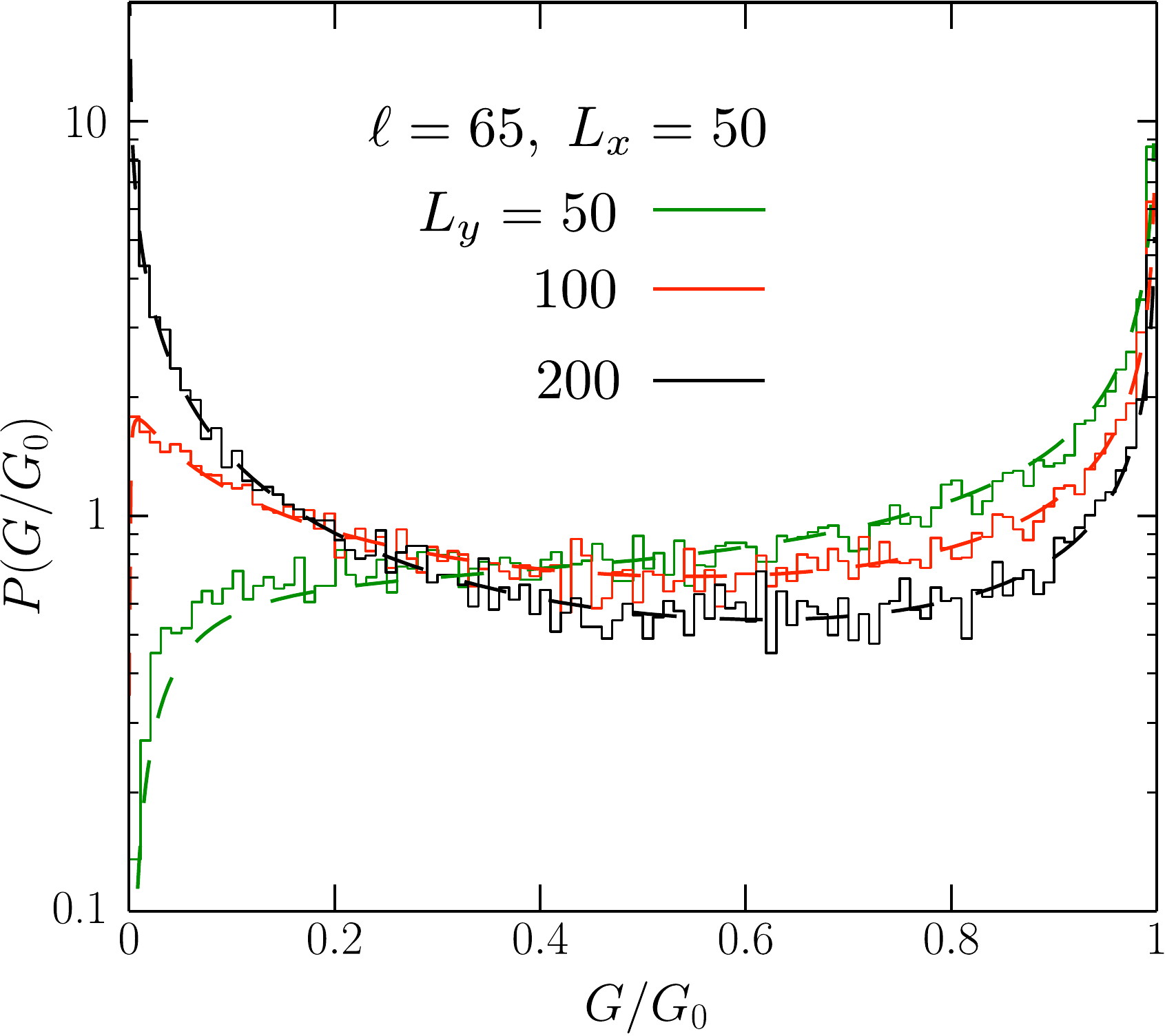}
  \end{center}
  \caption{Disorder-induced thermal conductance distribution of a single edge in the WTI phase of the anisotropic \textit{p}-wave superconductor in symmetry class D, for fixed $L_x=50$ and varying $L_y$. The histograms are calculated numerically ($\delta=t$, other parameters as in Fig.\ \ref{fig:DandDIIIscaling}). Dashed lines show the analytical result \eqref{PTbimodal} for $T\equiv G/G_0$, $L\equiv L_y$, with the mean free path $\ell=65$ as single fit parameter.}
  \label{fig:Gdist}
\end{figure}

We have tested the result \eqref{PTbimodal} in a computer simulation of the anisotropic \textit{p}-wave superconductor, with class-D Hamiltonian \eqref{eq:HclassDk}. In the geometry of Fig.\ \ref{fig_layout}\textit{b} both edges at $x=0$ and $x=L_x$ contribute to the thermal conductance in the WTI phase, but one edge can be removed by reducing the width of the contacts to the interval $0\leq x\leq L_x/2$. Results are shown in Fig.\ \ref{fig:Gdist}. With the mean free path $\ell$ as a single fit parameter, the transition from a uni-modal distribution to a bimodal distribution upon increasing $L/\ell$ is well-described by Eq.\ \eqref{PTbimodal}.

\section{Electrical detection of Kitaev edge modes}
\label{electrdetect}

So far we considered thermal conduction as the probe of edge state transport. Electrical detection would be more convenient experimentally, and this is possible by adapting the nanowire setup of Ref.\ \onlinecite{Akh11}. All contacts are now at the same temperature $T_0$, the superconductor is grounded as well as one of the metal contacts (number 2), and the other contact (number 1) is biased at voltage $V_1$. The electrical current into the grounded contact 2 fluctuates in time with noise power $P_{12}$. This is dominated by shot noise, at low temperatures when thermal noise can be neglected ($k_{\rm B}T_0\ll eV_1$). 

The noise power is given in terms of the transmission matrix by \cite{Ana96}
\begin{align}
P_{12}/P_0&={\rm Tr}\,\bigl(t_{ee}^{\dagger}t^{\vphantom{\dagger}}_{ee}+t_{he}^{\dagger}t^{\vphantom{\dagger}}_{he}\bigr)-{\rm Tr}\,\bigl(t_{ee}^{\dagger}t^{\vphantom{\dagger}}_{ee}-t_{he}^{\dagger}t^{\vphantom{\dagger}}_{he}\bigr)^{2}\nonumber\\
&=\tfrac{1}{2}\,{\rm Tr}\,t^{\dagger}t-\tfrac{1}{2}\,{\rm Tr}\,(\tau_{z}t^\dagger\tau_{z}t)^{2},\label{P12def}
\end{align}
with $P_0=e^{3}V_{1}/h$. The subscripts $e,h$ indicate transmission from electron to electron ($t_{ee}$) or from electron to hole ($t_{he}$), and we used electron-hole symmetry in the second equation to rewrite the whole expression in terms of the full transmission matrix $t$. 

\begin{figure}[tb]
  \begin{center}
	 \includegraphics[width=0.8\columnwidth]{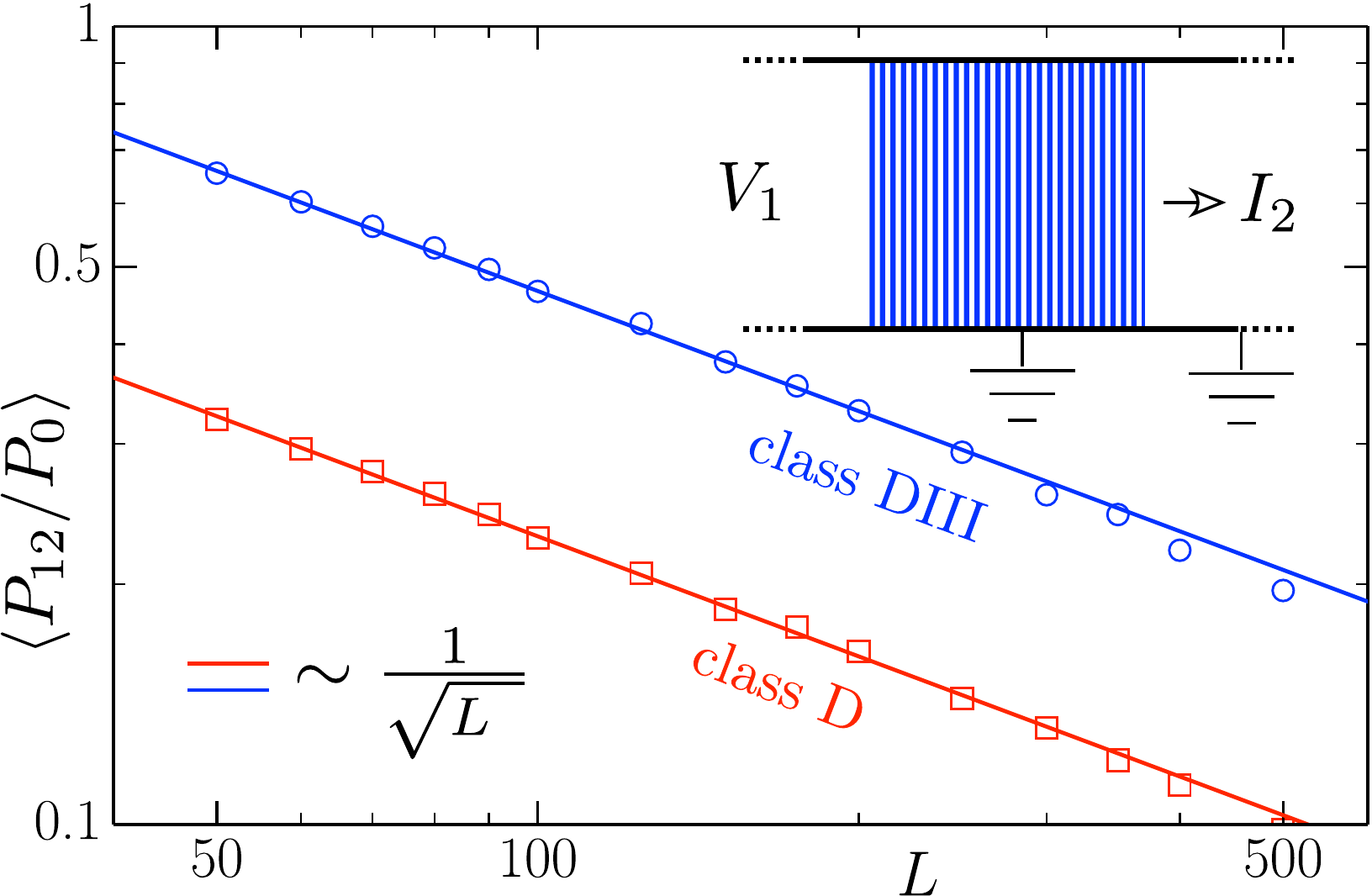}
  \end{center}
  \caption{Disorder-averaged electrical shot noise power \eqref{P12def} in the WTI phase of class D and DIII, for parameters as in Fig.\ \ref{fig:DandDIIIscaling}. Data points are averaged over $10^4$ disorder realizations. The solid lines show that this electrical transport property obeys the same $L^{-1/2}$ scaling as the thermal transport property of Fig.\ \ref{fig:DandDIIIscaling}. The inset shows the geometry, with one metal contact biased at voltage $V_1$ and both the superconductor and the second metal contact grounded. The electrical current $I_2$ into this second contact fluctuates in time with noise power $P_{12}=\int_{-\infty}^{\infty}dt\,\langle\delta I_2(0)\delta I_2(t)\rangle$.
  }
  \label{fig:PFscaling}
\end{figure}

As derived in Ref.\ \onlinecite{Akh11}, when the transmission is via an unpaired Majorana mode, the second trace in Eq.\ \eqref{P12def} vanishes identically so the electrical shot noise is directly related to the thermal conductance: $P_{12}/P_0=\frac{1}{2}G/G_{0}$. This applies to symmetry class D. More generally, in both symmetry classes D and DIII the two quantities $P_{12}$ and $G$ have the same $1/\sqrt{L}$ scaling in the WTI phase \cite{note2}, compare Figs.\ \ref{fig:DandDIIIscaling} and \ref{fig:PFscaling}.
 
\section{Conclusion}
\label{conclude}

The Kitaev model \cite{Kit01} is paradigmatic for topological superconductivity and Majorana zero-modes, and for that reason has been studied extensively \cite{Kit09,Ali12}. Here we have shown that the realization of this model at the edge of a two-dimensional system (what we have called the ``Kitaev edge'') is fundamentally different from its strictly one-dimensional counterpart, the Kitaev chain. The difference, summarized in Fig.\ \ref{fig_chainedge}, manifests itself in the different distribution of the thermal conductance, peaked at exponentially small value in the Kitaev chain \cite{Mot01,Bro03,Gru05} while the Kitaev edge has a second peak at the conductance quantum.

As a possible physical realization of Kitaev edge modes we have studied in some detail a model of an anisotropic two-dimensional chiral \textit{p}-wave superconductor \cite{Asa12,Ful12,Ser14,Buh14}, as well as its time-reversally symmetric (helical) counterpart. Both can produce weak topological insulators (WTI) with Kitaev edge modes, but while they appear at any amount of anisotropy for chiral \textit{p}-wave pairing, the helical \textit{p}-wave pairing requires a threshold anisotropy (compare Figs.\ \ref{fig:classDcleanphases} and \ref{fig:classDIIIcleanphases}). We have demonstrated the robustness of the WTI phase to disorder by numerical simulations, in good agreement with analytical calculations of the phase boundaries in self-consistent Born approximation (Figs.\ \ref{fig:phases_d_disorder} and \ref{fig:diii_phases_disorder}).

Experimentally the transition into the WTI phase can be detected, on length scales below the mean free path, via the doubling of the thermal conductance (Fig.\ \ref{fig:cleanEdgeG}), and on larger length scales via the super-Ohmic scaling (Fig.\ \ref{fig:DandDIIIscaling}). Because of the complexity of thermal transport measurements at low temperatures, we have proposed an alternative fully electrical method of detection, using the electrical shot noise power (Fig.\ \ref{fig:PFscaling}).

\acknowledgments

Discussions with A. R. Akhmerov are gratefully acknowledged. This research was supported by the Foundation for Fundamental Research on Matter (FOM), the Netherlands Organization for Scientific Research (NWO/OCW), and an ERC Synergy Grant.

\appendix
\section{Calculation of the phase boundaries in self-consistent Born approximation}
\label{scBa}

We calculate the phase diagram in the presence of electrostatic disorder (strength $\delta$ as defined in the main text)
using the self-consistent Born approximation (SCBA). 
Below we provide details of the calculation for the  class-DIII Hamiltonian \eqref{eq:HclassDIIIk}.
The corresponding results for the class-D Hamiltonian \eqref{eq:HclassDk} are simply obtained by taking the vanishing coupling $K \rightarrow 0$ limit
and are summarized at the end of this Appendix.

We calculate the disorder-averaged density of states from the self-energy $\Sigma$, defined by
\begin{align}
  &\frac{1}{E+i0^+-H_{\rm DIII}-\Sigma} = \nonumber\\
  & \qquad \left\langle \frac{1}{E + i0^+-H_{\rm DIII}-H_{\rm disorder}}\right\rangle.
  \label{eq:selfenergy}
\end{align}
The SCBA self-energy at the Fermi level ($E=0$) is given by
\begin{align}
  \Sigma = \tfrac{1}{3}\delta^2\sum_{\bm k} \tau_z \frac{1}{i0^+-H_{\rm DIII}(\bm k) -\Sigma} \tau_z .
  \label{eq:defselfscba}
\end{align}
The sum over $\bm k$ ranges over the first Brillouin zone and in the continuum limit
\begin{equation}
  \sum_{\bm k} \mapsto \frac{1}{4\pi^2}\int_{-\pi}^\pi dk_x\int_{-\pi}^\pi dk_y.
\end{equation}

The SCBA self-energy is a $\bm k$-independent $4\times 4$ matrix with spin and electron-hole degrees of freedom
\begin{align}
  \Sigma = (\sigma_0\otimes\tau_z)\delta \mu - (\sigma_y\otimes \tau_y) \delta K -(\sigma_0\otimes \tau_0)i\gamma .
  \label{eq:scbaselfenergy}
\end{align}
The terms $\delta\mu$ and $\delta K$ renormalize the chemical potential and coupling respectively. Both terms account for a disorder induced shift of the phase boundaries between the trivial insulator, TI, WTI, and thermal metal. The term $\gamma$ produces a finite density of states, induced by the disorder. Such a finite density of states may indicate a thermal metal or a trivial Anderson insulator, but it cannot distinguish between the two.

We substitute the self-energy~\eqref{eq:scbaselfenergy} into Eq.~\eqref{eq:defselfscba} and 
observe that the right-hand-side depends only on the renormalized chemical potential
$\tilde\mu=\mu-\delta\mu$ and coupling $\tilde K = K -\delta K$. Denoting the renormalized Hamiltonian
by $\tilde H_{\rm DIII}(\bm k)\equiv H_{\rm DIII}(\bm k) + (\sigma_0\otimes\tau_z)\delta \mu - (\sigma_y\otimes \tau_y) \delta K $
we can write the following identity:
\begin{align}
& \left(i\gamma - {\tilde H}_{\rm DIII}(\bm k) \right)^{-1} = \left(i\gamma + {\tilde H}_{\rm DIII}(\bm k) \right) \nonumber \\
  &\; \cdot\left(\gamma^2 +f_{\tilde \mu,\tilde K}(\bm k) - 2{\tilde K}(\Delta_x\sin k_x\, \sigma_z \tau_z + \Delta_y\sin k_y\, \sigma_y  )\right) \nonumber \\
  &\; \cdot \left( 4 {\tilde K}^2 g(\bm k) - \left(\gamma^2 + f_{\tilde \mu,\tilde K}(\bm k)\right)^2  \right)^{-1},
\end{align}
where
\begin{subequations}
\begin{align}
 g(\bm k) &= \Delta_x^2\sin^2 k_x + \Delta_y^2\sin^2 k_y,\\
  f_{\tilde \mu, \tilde K}(\bm k) &=  \epsilon_{\tilde \mu}(\bm k)^2 + {\tilde K}^2 + g(\bm k).
  \label{eq:fg}
\end{align}
\end{subequations}
Because of symmetry only terms even in $\bm k$ contribute to $\sum_{\bm k}$. 
We find from Eq.~\eqref{eq:defselfscba} three coupled equations for the parameters $\delta \mu$, $\delta K$, and $\gamma$, which completely determine the SCBA self-energy:
\begin{subequations}
\begin{align}
  \delta\mu &= \tfrac{1}{3}\delta^2\sum_{\bm k} \frac{ \epsilon_{\tilde \mu}(\bm k) (\gamma^2 + f_{\tilde\mu, \tilde K}(\bm k) )}
{ 4{\tilde K}^2g(\bm k) -
(\gamma^2 + f_{\tilde \mu, \tilde K}(\bm k))^2  }, \\
  \delta K &= \tfrac{1}{3}\delta^2\sum_{\bm k} \tilde K \frac{ \gamma^2 + f_{\tilde\mu, \tilde K}(\bm k) -2 g(\bm k) }
{ 4{\tilde K}^2g(\bm k) -
(\gamma^2 + f_{\tilde \mu, \tilde K}(\bm k))^2  }, \\
  \gamma &= \tfrac{1}{3}\delta^2\sum_{\bm k} \frac{-\gamma (\gamma^2 + f_{\tilde\mu, \tilde K}(\bm k) )}
{ 4{\tilde K}^2g(\bm k) - (\gamma^2 + f_{\tilde \mu, \tilde K}(\bm k))^2  }.
\end{align}
  \label{eq:deltamu_deltaK_gamma}
\end{subequations}

We address first the shift of phase boundaries in the weak disorder case.
To this end we set $\gamma=0$ assuming that the disorder is too weak to induce a finite density of states. 
We are looking for solutions of the clean system gap closing conditions Eqs.~\eqref{eq:mu_ti}, \eqref{eq:mu_sti}, \eqref{eq:mu_triv} expressed in terms of the renormalized parameters $\mu\rightarrow\tilde\mu$, $K\rightarrow\tilde K$. The gap closing condition defines the phase boundary and can be expressed as $\tilde K = \tilde K(\tilde\mu)$ (with a different function $\tilde K(\tilde \mu)$ for each boundary).
We rewrite the SCBA equations \eqref{eq:deltamu_deltaK_gamma} with $\gamma=0$ in the form
\begin{subequations}
\begin{align}
  \mu &=  
         \tfrac{1}{3}\delta^2 F(\tilde\mu, \tilde K) + \tilde\mu,\label{eq:mu_mutilde}\\
\delta^2 &= 3(K-\tilde K)\left(G(\tilde\mu, \tilde K)\right)^{-1},
\end{align}
\end{subequations}
where
\begin{align}
  F(\tilde\mu, \tilde K)  &=  
          \sum_{\bm k} \frac{\epsilon_{\tilde \mu}(\bm k) f_{\tilde\mu, \tilde K}(\bm k)}
{ 4 {\tilde K}^2 g(\bm k) - f_{\tilde \mu, \tilde K}^2(\bm k)},\\
G(\tilde\mu, \tilde K)    &= 
          \sum_{\bm k} {\tilde K} \frac{ f_{\tilde\mu, \tilde K}(\bm k) - 2 g(\bm k) }
{ 4 {\tilde K}^2 g(\bm k) - f_{\tilde \mu, \tilde K}^2(\bm k)}.
\end{align}

In  this way we obtain the parametric solution for the disorder strength $\delta^2[\tilde\mu, \tilde K(\tilde\mu)]$ and the gap parameter $\mu[\tilde\mu, \delta^2(\tilde\mu)]$ along the phase boundary. We vary the parameter
$\tilde\mu$ away from the clean system solution $\mu$ at $\delta=0$. The sums over the Brillouin zone are computed numerically in the continuum limit. The resulting parametric phase boundaries $[\mu(\tilde\mu),\delta(\tilde\mu)]$ separate insulating and gapless phases at low to moderate disorder.

For sufficiently large disorder $\delta>\delta_c$ the SCBA equations \eqref{eq:deltamu_deltaK_gamma} may support solutions with non-zero $\gamma$ indicating the onset of a finite density of states at zero energy.
This marks the transitions from the strong and weak topological insulators to the thermal metal at strong disorder.
The $\delta_c$ dependence  on $\mu$ and $K$ follows from solutions of the SCBA equations at infinitesimal $\gamma\neq 0$
\begin{subequations}
\begin{align}
  1 &= \frac{\delta_c^2}{3}\sum_{\bm k} \frac{f_{\tilde\mu, \tilde K}(\bm k)}{f^2_{\tilde \mu, \tilde K}(\bm k) - 4 {\tilde K}^2 g(\bm k)} \equiv \frac{\delta_c^2}{3} H(\tilde\mu, \tilde K), \label{eq:delta_c}\\
  \tilde\mu &= - \tfrac{1}{3} \delta^2_c F(\tilde\mu, \tilde K) + \mu, \label{eq:mu_tilde_of_delta_c}\\
  \tilde K  &= - \tfrac{1}{3} \delta^2_c G(\tilde\mu, \tilde K) + K. \label{eq:K_tilde_of_delta_c}
\end{align}
\end{subequations}
To determine $\delta_c$ we first search numerically (using Steffensen iteration) for fixed point solutions of Eqs.~\eqref{eq:mu_tilde_of_delta_c},~\eqref{eq:K_tilde_of_delta_c},
\begin{align}
  \begin{pmatrix}
	 \tilde\mu \\ \tilde K
  \end{pmatrix} = \frac{-1}{H(\tilde\mu,\tilde K)}
  \begin{pmatrix}
	 F(\tilde\mu, \tilde K) \\ G(\tilde\mu, \tilde K)
  \end{pmatrix} +
  \begin{pmatrix}
	 \mu \\ K
  \end{pmatrix},
  \label{eq:fixpoint}
\end{align}
for a given value of the chemical potential $\mu$ and coupling $K$.
Finally we compute $\delta_c$ from \eqref{eq:delta_c} for the obtained solutions $(\tilde\mu, \tilde K)$.

Both the parametric solutions to the renormalized gap closing conditions and the computed $\mu$-dependence (for fixed $K$) of the critical disorder $\delta_c$ are shown as white dashed lines in Fig.~\ref{fig:diii_phases_disorder}.

This was all for class DIII. The formulas for class D correspond to the $K\rightarrow 0$ limit.
Denoting by $E_{\mu}(\bm k)$ the excitation spectrum of the class-D Hamiltonian \eqref{eq:HclassDk} we find
\begin{align}
E^2_\mu(\bm k) & =  f_{\mu, K\rightarrow 0}(\bm k) = \epsilon^2_{\mu}(\bm k) + \Delta_x^2\sin^2 k_x + \Delta_y^2\sin^2 k_y.
  \label{eq:Emuk}
\end{align}
The shift of the phase boundaries at low disorder is obtained by imposing the gap closing conditions for the renormalized chemical potential $\tilde\mu=\pm2t(1\pm\alpha)$.
From Eq.~\eqref{eq:mu_mutilde} we directly obtain the boundary position as a function of the disorder strength
\begin{align}
  \mu = \left.\left[-\tfrac{1}{3}\delta^2\sum_{\bm k} \frac{\epsilon_{\tilde \mu}(\bm k)}{E^2_{\tilde \mu}(\bm k)} + \tilde\mu \right]\right|_{\tilde \mu=\pm 2t(1\pm\alpha)}.
\end{align}
For the isotropic case $\alpha=1$ the central $\mu=0$ transition separating the two TI phases is not renormalized by disorder. The transitions separating the TI phases from the trivial insulator phase are shifted according to $\mu = \pm( 4t + 0.130\,\delta^2)$.
For $\alpha=1/2$ the transition line between the WTI and TI phases is given by $\mu = \pm(3t+ 0.039\,\delta^2)$ and the transition between TI and the trivial phase by $\mu = \pm (t + 0.180\,\delta^2)$. These phase boundaries are shown as dashed lines in Fig.~\ref{fig:phases_d_disorder}.

The critical disorder lines separating the TI and WTI phases from the thermal metal at large disorder can be found from the SCBA solutions Eqs.~\eqref{eq:delta_c},~\eqref{eq:mu_tilde_of_delta_c} at infinitesimal $\gamma\neq 0$.
In class-D we can directly parametrize such solutions by the renormalized chemical potential $\tilde\mu$.
We obtain
\begin{align}
  \delta_c^2(\tilde\mu) &= 3 \left( \sum_{\bm k}\frac{1}{E^2_{\tilde\mu}(\bm k)} \right) ^{-1}, \\
  \mu\left[\tilde \mu, \delta_c(\tilde\mu)\right] &= \tilde\mu - \tfrac{1}{3} \delta_c^2(\tilde\mu)\sum_{\bm k}\frac{\epsilon_{\tilde\mu}(\bm k)}{E_{\tilde\mu}^2(\bm k)}.
  \label{eq:deltac_mu_of_tildemu}
\end{align}
The parametric dependence for $\tilde \mu \in [-4,4]$ calculated from the above equations is also included in Fig.~\ref{fig:phases_d_disorder}.


\begin{thebibliography}{99}
\bibitem{Hal82} B. I. Halperin, Phys. Rev. B \textbf{25}, 2185 (1982).
\bibitem{But88} M. B\"{u}ttiker, Phys. Rev. B \textbf{38}, 9375 (1988).
\bibitem{Has10} M. Z. Hasan and C. L. Kane, Rev. Mod. Phys. \textbf{82}, 3045 (2010).
\bibitem{Qi11} X.-L. Qi and S.-C. Zhang, Rev. Mod. Phys. \textbf{83}, 1057 (2011).
\bibitem{Asa12} D. Asahi and N. Nagaosa, Phys. Rev. B \textbf{86}, 100504(R) (2012).
\bibitem{Ful12} I. C. Fulga, B. van Heck, J. M. Edge, and A. R. Akhmerov, arXiv:1212.6191.
\bibitem{Ser14} I. Seroussi, E. Berg, and Y. Oreg, arXiv:1401.2671.
\bibitem{Buh14} A. B\"{u}hler, N. Lang, C. V. Kraus, G. M\"{o}ller, S. D. Huber, and H. P. B\"{u}chler, arXiv:1403.0593.
\bibitem{Kit01} A. Yu. Kitaev, Phys.-Usp. Suppl. \textbf{44}, 131 (2001).
\bibitem{Bro00} P. W. Brouwer, A. Furusaki, I. A. Gruzberg, and C. Mudry, Phys. Rev. Lett. \textbf{85}, 1064 (2000).
\bibitem{Akh11} A. R. Akhmerov, J. P. Dahlhaus, F. Hassler, M. Wimmer, and C. W. J. Beenakker, Phys. Rev. Lett. \textbf{106}, 057001 (2011).
\bibitem{Mot01} O. Motrunich, K. Damle, and D. A. Huse, Phys. Rev. B \textbf{63}, 224204 (2001).
\bibitem{Bro03} P. W. Brouwer, A. Furusaki, and C. Mudry, Phys. Rev. B \textbf{67}, 014530 (2003); P. W. Brouwer, A. Furusaki, C. Mudry, and S. Ryu, arXiv:cond-mat/0511622. 
\bibitem{Gru05} I. A. Gruzberg, N. Read, and S. Vishveshwara, Phys. Rev. B \textbf{71}, 245124 (2005).
\bibitem{Cho11} T.-P. Choy, J. M. Edge, A. R. Akhmerov, and C. W. J. Beenakker, Phys. Rev. B \textbf{84}, 195442 (2011).
\bibitem{Nad13} S. Nadj-Perge, I. K. Drozdov, B. A. Bernevig, and A. Yazdani, Phys. Rev. B \textbf{88}, 020407 (2013).
\bibitem{Alt97} A. Altland and M. R. Zirnbauer, Phys. Rev. B \textbf{55}, 1142 (1997).
\bibitem{kwant} C. W. Groth, M. Wimmer, A. R. Akhmerov, and X. Waintal, arXiv:1309.2926.
\bibitem{Tho82} D. J. Thouless, M. Kohmoto, M. P. Nightingale, and M. den Nijs, Phys. Rev. Lett. \textbf{49}, 405 (1982).
\bibitem{Bra10} G. Br\"{a}unlich, G. M. Graf, and G. Ortelli, Commun. Math. Phys. {\bf 295}, 243 (2010).
\bibitem{Ful12c} I. C. Fulga, F. Hassler, and A. R. Akhmerov, Phys. Rev. B \textbf{85}, 165409 (2012).
\bibitem{note1} When we implement the twisted periodic boundary condition on the lattice it should extend over an odd number of sites, in order to avoid a minigap in the edge state spectrum that would spoil the calculation of the weak topological invariant (Figs.\ \ref{fig:dQ} and \ref{fig:diiiQ}). For the conductance it makes no difference whether there is an even or an odd number of lattice sites across.
\bibitem{Fu07} L. Fu and C. L. Kane, Phys. Rev. B \textbf{76}, 045302 (2007).
\bibitem{Fu07b} L. Fu, C. Kane, and E. Mele, Phys. Rev. Lett. \textbf{98}, 106803 (2007).
\bibitem{Sen00} T. Senthil and M. P. A. Fisher, Phys. Rev. B \textbf{61}, 9690 (2000).
\bibitem{Eve08} F. Evers and A. D. Mirlin, Rev. Mod. Phys. \textbf{80}, 1355 (2008).
\bibitem{Med10} M. V. Medvedyeva, J. Tworzyd{\l}o, and C. W. J. Beenakker, Phys. Rev. B \textbf{81}, 214203 (2010).
\bibitem{Ful12b} I. C. Fulga, A. R. Akhmerov, J. Tworzyd{\l}o, B. B{\'{e}}ri, and C. W. J. Beenakker, Phys. Rev. B \textbf{86}, 054505 (2012).
\bibitem{Ful11} I. C. Fulga, F. Hassler, A. R. Akhmerov, and C. W. J. Beenakker, Phys. Rev. B \textbf{83}, 155429 (2011).
\bibitem{Wan14} D. Wang, Z. Huang, and C. Wu, arXiv:1401.3323.
\bibitem{Wak14} R. Wakatsuki, M. Ezawa, and N. Nagaosa, arXiv:1401.5192.
\bibitem{Dum14} E. Dumitrescu, T. D. Stanescu, and S. Tewari, arXiv:1403.3093.
\bibitem{Ana96} M. P. Anantram and S. Datta, Phys. Rev. B \textbf{53}, 16390 (1996).
\bibitem{note2} The first trace in Eq.\ \eqref{P12def} is proportional to the transmission probability $T$, while the second trace is proportional to $T^2$. Because of the bimodal distribution \eqref{PTbimodal}, the averages of $T$ and $T^2$ scale with the same power of $L$: $\langle T^2\rangle=\frac{2}{3}\langle T\rangle\propto \sqrt{1/L}$. Incidentally, we note that this also implies that the ``thermal Fano factor'' $F=\langle T(1-T)\rangle\langle T\rangle^{-1}$ has the same $L$-independent value $1/3$ as the electrical Fano factor of a metallic diffusive conductor --- even though the conductance in that system has a Gaussian rather than bimodal distribution.
\bibitem{Kit09} A. Kitaev, in: \textit{Topological phases and quantum computations: Exact Methods in Low-Dimensional Statistical Physics and Quantum Computing}, edited by J. Jacobsen et al. (Oxford University Press, 2010); A. Kitaev and C. Laumann, arXiv:0904.2771.
\bibitem{Ali12} J. Alicea, Rep. Prog. Phys. \textbf{75}, 076501 (2012).
\end{thebibliography}
\end{document}